\documentclass[%
 aip,
 sd,%
 amsmath,amssymb,
preprint,%
groupedaddress
]{revtex4-1}
\usepackage{mathbbol}
\usepackage{graphicx}
\usepackage{dcolumn}
\usepackage[
breaklinks,
colorlinks,
linkcolor=blue,
citecolor=blue,
urlcolor=blue]{hyperref}

\usepackage{color,soul}

\begin{document}

\title{A short trajectory is all you need: A transformer-based model for long-time dissipative quantum dynamics}

\author{Luis E. Herrera Rodríguez}
\author{Alexei A. Kananenka}%
 \email{akanane@udel.edu}
\affiliation{%
 Department of Physics and Astronomy, University of Delaware, Newark, DE 19716, USA
}%

\date{\today}

\begin{abstract}
In this communication we demonstrate that a deep artificial neural network based on a transformer architecture with self-attention layers can predict the long-time population dynamics of a quantum system coupled to a dissipative environment provided
that the short-time population dynamics of the system is known. The transformer neural network model developed
in this work predicts the long-time dynamics of  spin-boson model efficiently and very accurately across different regimes, from weak system-bath coupling  to strong coupling non-Markovian regimes. Our model is more accurate than classical forecasting models, such as recurrent neural networks and is comparable to the state-of-the-art models for simulating the dynamics of quantum dissipative systems based on kernel ridge regression. 

\end{abstract}

\maketitle



Simulating long-time nonadiabatic dynamics of  
 realistic complex molecular or condensed-phase systems
comprised of a large number of, generally, anharmonic degrees of freedom (DOFs) remains a challenge.
In the condensed-phase
the thermal reservoir or bath  induces an irreversible dissipation to the system altering its dynamics  from unitary to non-unitary.\cite{weiss2012quantum,breuer2002theory} A plethora of numerically exact methods were developed to attack this problem including hierarchical equations of motion (HEOM), \cite{tanimura1989two,tanimura2020numerically} multi-configurational time-dependent Hartree, \cite{wang2003multilayer,meyer1990multi} quasi-adiabatic propagator path integral,\cite{kundu2023pathsum, makarov1994path}  time-dependent Davydov ansatz,\cite{luo2010validity} time-dependent density matrix renormalization group, \cite{ren2018time} tensor-train split-operator Fourier transform,\cite{greene2017tensor} and
  the stochastic equation of motion approach \cite{yan2016stochastic,hsieh2018unified,hsieh2018unified,han2020stochastic,ullah2020stochastic} to name just a few. These methods, in general, are restricted to specific forms of the Hamiltonian (e.g., harmonic bath) and have a computational cost that makes long-time simulations, even with the model Hamiltonians, often infeasible.  Projection-based methods like the Nakajima--Zwanzig generalized quantum master equation (GQME) allow to simulate long-time quantum dynamics at a lower computational cost, provided the memory kernel is known \textit{a priori}. \cite{brian2021generalized, nakajima1958quantum,zwanzig1960ensemble,shi2003new,mulvihill2021road,kelly2013efficient} However, calculating the numerically exact memory kernel for GQME for a general multi-level quantum system coupled to a bath is extremely difficult. Transfer tensor method  reduces the computational cost compared with GQME but it requires a set of short-time system's reduced density matrices as an input which 
must be generated by a numerically accurate method.\cite{cerrillo2014non,kananenka2016accurate,buser2017initial,gelzinis2017applicability,chen2020non} Therefore, it possesses the same limitations as the abovementioned methods with the exception that the computationally expensive simulations are needed only to generate the short-time input.  

In recent years several machine-learning (ML) based approaches for simulating long-time dynamics of quantum dissipative systems have been developed.\cite{herrera2021convolutional,ullah2021speeding,ullah2022predicting,ullah2022one,ullah2023qd3set,ullah2024mlqd,lin2021simulation,wu21,akimov21,zhang24,lin22,luo2009autoregressive,rodriguez22,D4DD00153B} Such methods dramatically reduce the computational cost of quantum dynamics simulations while, in many cases, maintain high accuracy and systematic improvability achieved by increasing the number of trainable parameters and/or using larger data sets for model training. Several types of ML models have been explored to date including feed-forward fully-connected neural networks (FFNN),\cite{akimov21,rodriguez22} convolutional neural networks, \cite{herrera2021convolutional,ullah2022predicting,ullah2022one,wu21,rodriguez22} (bidirectional) recurrent neural networks (RNN), convolutional recurrent neural networks, \cite{lin2021simulation,lin22,rodriguez22} and kernel ridge regression (KRR). \cite{ullah2021speeding,ullah2024mlqd,zhang24,rodriguez22}
For example, Rodríguez \textit{et al.}\cite{rodriguez22} showed that KRR models outperform neural-network-based (NN) models, for predicting the long-time dynamics. However, in the study of Rodríguez \textit{et al.}\cite{rodriguez22} the data set size was restricted due to poor scaling of KRR methods with the size of the training set.

In this Communication we present a neural network model, based on the transformer architecture, for simulating
the long-time population dynamics of a quantum system coupled to the bath.
The transformer model was originally introduced by Vaswani \textit{et al.}\cite{vaswani2017attention} in the context of natural language processing (NLP) and it quickly became the state of the art for NLP. The transformer neural network  learns long-range relationships or contextual information from  sequential data.  This is in contrast to RNNs where the range is limited by the ``memory''. The transformer architecture has become popular not only in NLP but also in computer vision and audio.\cite{dosovitskiy2020image,radford2023robust} Additionally, recently developed pre-trained models such as GPT (generative pre-trained transformers) and BERT (bidirectional encoder representations from transformers) are based on the transformer architecture.\cite{wolf2020transformers} Recognizing that transformers are effectively general purpose trainable computers whose domain of applicability is not limited to NLP tasks, they
are now being recruited to address various problems in computational physics.
 Recently the transformer architecture was used to classify light curves of astronomical objects (Astromer).\cite{donoso2023astromer} 
 In the present work we optimized and trained a transformer model to predict the long-time dynamics of a quantum dissipative system.  To the best of our knowledge, this is the first application of the transformer architecture in the context of modeling the time-evolution of open quantum systems.



We choose to test our transformer implementation on a time series representing the reduced density matrix dynamics of the spin-boson model which is a cornerstone model in the study of open quantum systems due to its rich physics and widespread applicability.\cite{leggett1987dynamics} The applications range from quantum computing,\cite{makhlin2001quantum} quantum phase transitions, \cite{winter2008quantum,alvermann2008sparse} to electron transfer in biological systems.\cite{garg1985effect} The spin-boson model comprises a two-level quantum subsystem linearly coupled to a bosonic bath environment which is modeled as an ensemble of independent quantum harmonic oscillators. The total Hamiltonian in the subsystem's basis denoted as  $\{|0\rangle,|1\rangle\}$ is given by $(\hbar=1)$
\begin{equation}
    \hat{H}=\epsilon \hat{\sigma}_z+\Delta \hat{\sigma}_x+\hat{\sigma}_z \sum_\alpha g_\alpha\left(\hat{b}_\alpha^{\dagger}+\hat{b}_\alpha\right)+\sum_\alpha \omega_\alpha \hat{b}_\alpha^{\dagger} \hat{b}_\alpha,\label{eq:ham}
\end{equation}
where  $\hat{\sigma}_z=|0\rangle\langle0|-|1\rangle\langle 1|$ and  $\hat{\sigma}_x=|0\rangle\langle1|+|1\rangle\langle0|$ are the Pauli operators, $\hat{b}_\alpha^{\dagger}\left(\hat{b}_\alpha\right)$ is the bosonic creation (annihilation) operator of the $\alpha$th mode with the frequency $\omega_\alpha$, $\epsilon$ is the energetic bias, $\Delta$ is the tunneling matrix element, and $g_\alpha$ are the subsystem-bath coupling coefficients.

The description of the bath is completely determined by the spectral density $J(\omega)=\pi \sum_\alpha g_\alpha^2 \delta\left(\omega_\alpha-\omega\right)$ which, in this work, is chosen to be of the Debye form (Ohmic spectral density with the Drude-Lorentz cut-off) \cite{wang1999semiclassical}
\begin{equation}
    J(\omega)=2 \lambda \frac{\omega \omega_c}{\omega^2+\omega_c^2},
\end{equation}
where $\lambda$ is the bath reorganization energy which controls the strength of system-bath coupling and $\omega_c$ is the cutoff frequency.  
We focus on the time evolution of the expectation value of $\hat{\sigma}_z$ Pauli operator
\begin{equation}
    \left\langle\hat{\sigma}_z(t)\right\rangle=\operatorname{Tr}_{\mathrm{s}}\left[\hat{\sigma}_z \hat{\rho}_{\mathrm{s}}(t)\right],
    \label{eq:expecte_value}
\end{equation}
which is often referred to as the population difference $\left\langle\hat{\sigma}_z(t)\right\rangle=p_0(t)-p_1(t)$, where  $p_{ 0}(t)=\operatorname{Tr}_{\mathrm{s}}$ $\left[| 0\rangle\langle 0| \hat{\rho}_s(t)\right]$ and $p_{ 1}(t)=\operatorname{Tr}_{\mathrm{s}}$ $\left[| 1\rangle\langle 1| \hat{\rho}_s(t)\right]$. In Eq.~\eqref{eq:expecte_value} the trace is taken over the subsystem's DOFs as denoted by 's' and $\hat{\rho}_{\mathrm{s}}$ is the subsystem's reduced density operator
\begin{equation}
    \hat{\rho}_{\mathrm{s}}(t)=\operatorname{Tr}_{\mathrm{b}}\left[e^{-i \hat{H} t} \hat{\rho}(0) e^{i \hat{H} t}\right],
\end{equation}
where $\hat{\rho}(0)$ is the total system plus bath density operator and the trace is taken over the bath DOFs. The initial state of the total system is assumed to be the product state of the following form
\begin{equation}
    \hat{\rho}(0)=\hat{\rho}_{\mathrm{s}}(0) \otimes \frac{e^{-\beta \hat{H}_{\mathrm{b}}}}{Z_{\mathrm{b}}},
\end{equation}
where $\hat{H}_b=\sum_\alpha \omega_\alpha b_\alpha^{\dagger} b_\alpha$ is the bath Hamiltonian, $Z_{\mathrm{b}}=\operatorname{Tr}_{\mathrm{b}}\left[e^{-\beta \hat{H}_{\mathrm{b}}}\right]$ is the bath partition function, $\beta=\left(k_{\mathrm{B}} T\right)^{-1}$ is the inverse temperature, and $k_{\mathrm{B}}$ is the Boltzmann constant. The initial density operator of the subsystem is chosen to be $\hat{\rho}_{\mathrm{s}}(0)=|0\rangle\langle 0|$. This initial condition corresponds to situations where the initial preparation of the subsystem occurs quickly on a timescale of the bath relaxation.

 In this work the $i$th input into the ML model is given by a pair of $\{\mathbf{x}_i,\mathbf{t}_i\}$ where $\mathbf{x}_i=\left(x^{(1)}_i, \ldots, x^{(T)}_i\right)$ is a time-ordered sequence of  expectation values of $\hat{\sigma}_z (t)$  $x^{(j)}=\langle\hat{\sigma}_z (t_j)\rangle$ (the population difference)  and $\mathbf{t}_i$ is the vector of the corresponding times $\mathbf{t}_i=\left(t^{(1)}_i, \ldots, t^{(T)}_i\right)$ where $T$ is the length of the input time series. Each element of the input vectors is a pair of real-valued numbers $x^{(j)} ,  t^{(j)} \in \mathbb{R}$. Consider a data set $\mathcal{D}=\left\{\left(\{\mathbf{x}_i, \mathbf{t}_i \} , \mathbf{y}_i\right)\right\}_{i=1}^N$ containing $N$ time series $\mathbf{x}_i$, the corresponding times $\mathbf{t}_i$  and their associated labels $\mathbf{y}_i$. In a time-series forecasting problem, the labels can describe the future states of the input sequence $\mathbf{x}_i$ as denoted by $\mathbf{y}_i=\left(x_i^{(T+1)}, \ldots, x_i^{(T+m)}\right)$. In this work we train a transformer-based NN model to predict a single real-valued scalar quantity, the population difference of the spin-boson model $\left\langle\hat{\sigma}_z(t)\right\rangle$ for a single time step, $m=1$, following the last input timestep, $T$. Extensions to multi-step outputs $(m>1)$ within the presented framework are straightforward. 
 

 The two main components of the transformer-based model are the self-attention mechanism and the positional encoding.
 
\textit{Self attention block}. Attention is one of the most crucial concepts in deep learning.\cite{niu2021review} It was inspired by the processing of large data. Humans tend to focus on distinctive parts of the information rather than processing it as a whole.\cite{corbetta2002control} In deep learning, especially in NLP, the attention mechanisms grant the models the ability to focus on specific parts of the input that have more relevance when producing the output. In other words, each element of the output is conditioned on the selection of items in the input of the model. 

The classical attention mechanism in NLP is 
based on RNN and is called attention-based RNN. The model consists of an RNN that encodes the input and an RNN that decodes it. In between, the attention model is added which allows the attention-based RNN to focus on the parts of the input that are critical for predicting the output target.\cite{bahdanau2014neural,aliabadi2020attention} 
Vaswani \textit{et al.} introduced the self-attention mechanism which, in contrast to the classical attention-based RNN, quantifies the importance of the relationships between parts of the input without conditioning it to the sequential order.\cite{vaswani2017attention} This allows self-attention models to be trained in parallel which improves the efficiency.


\begin{figure}[ht!]
    \centering
    \includegraphics[width=\columnwidth]{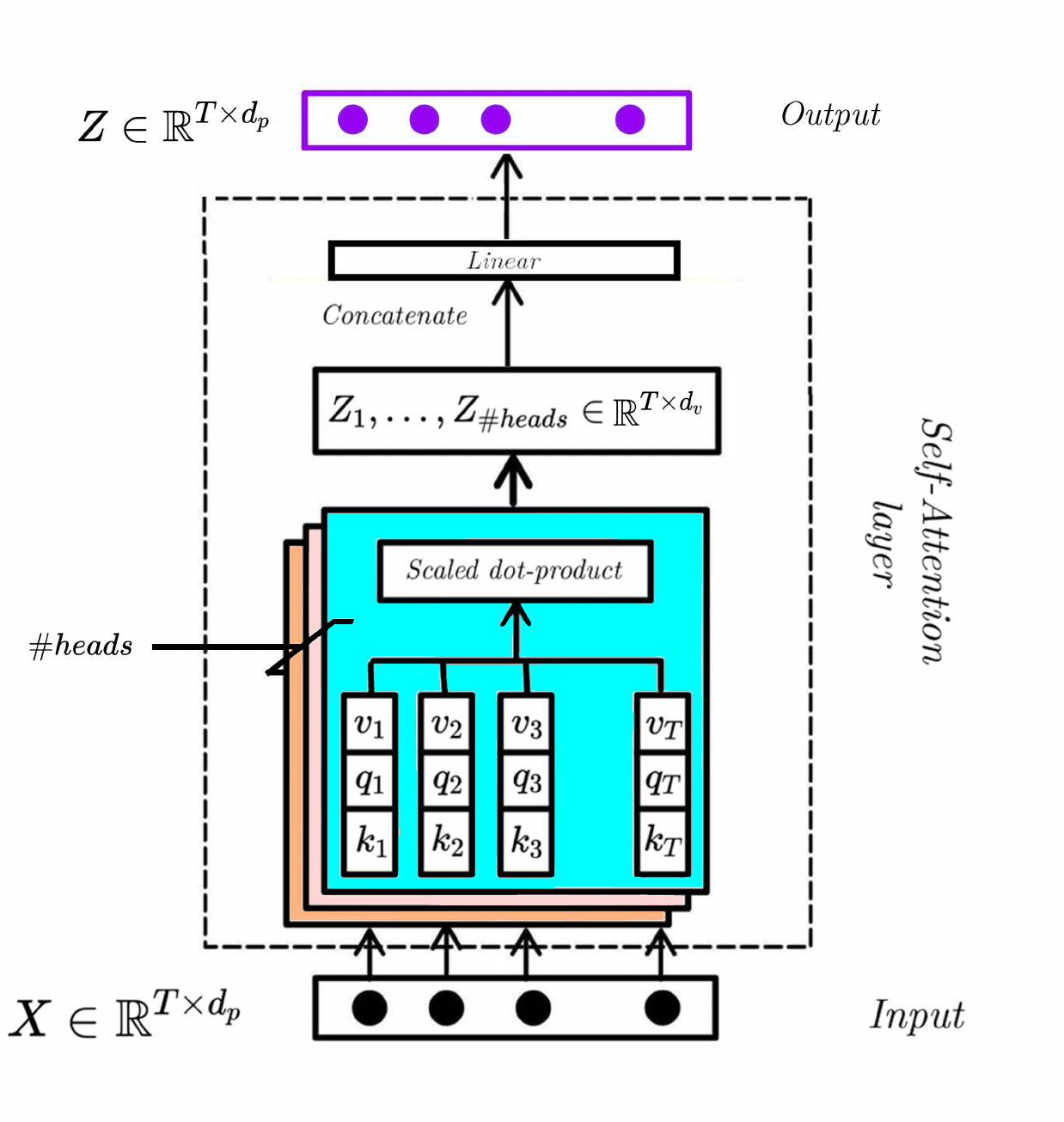}
    \caption{Self-attention layer diagram. Each input vector $X^{(j)} \in \mathbb{R}^{d_p}$ of a sequence, represented by solid black circles, is projected into the  (k)ey, (q)uery, and (v)alue vectors. The scaled dot product is calculated according to Eq. \eqref{eq:similarity} for each head to obtain output $Z_i$. There are three heads on this diagram, each is represented by a colored box. Finally, all the outputs are concatenated and projected into the output representation $Z \in \mathbb{R}^{T \times d_p}$ (solid purple circles). }
    \label{fig:SA-L}
\end{figure}

A self-attention layer consists of multiple self-attention heads, that compute the cosine similarities of every part of the input with itself and with any other part of the sequence.
Cosine similarity  is a measure of similarity between two non-zero vectors $\mathbf{v}$ and $\mathbf{u}$ defined as $S(\mathbf{v},\mathbf{u})=\mathbf{v}\cdot\mathbf{u} ||\mathbf{v}||^{-1}||\mathbf{u}||^{-1}$, where $||\mathbf{u}||$ is a norm of vector $\mathbf{u}$.
Thus, every self-attention head measures the relationship between pairs of elements of the sequence. If one element of the sequence affects the other, more attention is taken into account by the model. Each self-attention head represents the data in a different form, in this way  
different heads can capture different correlations. The self-attention layer receives as input a vector representation  $X \in \mathbb{R}^{T \times d_p} $(see \textit{Positional Encoding} below)  of the input sequence and transforms the input data into $Z \in \mathbb{R}^{T \times d_p} $ representation that is meaningful for the model. 
The self-attention matrix $Z_i \in \mathbb{R}^{T \times d_v} $ for the $i$th head is a weighted sum over the input values $\mathrm{V}_i$ in a query-key fashion

\begin{equation}
    \mathrm{Z}_i=\operatorname{softmax}\left(\frac{\mathrm{Q}_i \mathrm{~K}_i^{\top}}{\sqrt{\mathrm{d}_{\mathrm{k}}}}\right) \mathrm{V}_i \text {, }
    \label{eq:similarity}
\end{equation}
where the queries, keys, and values $\left(\mathrm{Q}_i, \mathrm{~K}_i, \mathrm{~V}_i\right)$ are  learnable input  transformations or projections
\begin{equation}
\mathrm{Q}_i=\mathrm{XW}_i^{\mathrm{q}}, \quad \mathrm{K}_i=\mathrm{XW}_i^{\mathrm{k}} \text {, and } \mathrm{V}_i=\mathrm{XW}_i^{\mathrm{v}} \text {, }
\end{equation}
where $\mathrm{W}_i^{\mathrm{q}} \in \mathbb{R}^{d_p \times d_k}, \mathrm{W}_i^{\mathrm{k}} \in \mathbb{R}^{d_p \times d_k}$, and $\mathrm{W}_i^{\mathrm{v}} \in \mathbb{R}^{d_p \times d_v}$ are trainable weight matrices of the $i$th head, $d_k $  specify the embedding size of each self-attention head and $d_v$ specifies the output dimension. 
The final output of the self-attention layer is the concatenation of the output of every single head and a projection $ Z = \text{CONCAT}\left\{\mathrm{Z}_1, \ldots, \mathrm{Z}_{\mathrm{i}}, \ldots, \mathrm{Z}_{\# \text { heads }}\right\} W^o$, with $W^o \in \mathbb{R}^{\# \text{heads} \; d_v \times d_p}$ as shown in the Fig.~\ref{fig:SA-L} where '$\#\text{Heads}$' is the number of heads. In this work we choose $d_k = d_p $ and $d_v = d_p/ \#\text{Heads}. $ 

\textit{Positional Encoding}. A set of expectation values $\langle \sigma _z (t_j) \rangle$ is clearly an important input information for the given problem. When using the transformer model the values of corresponding times are important as well, and is crucial when the input data is not equally spaced (which is not the case here but can be considered as a possible extension of this work). The self-attention layer in principle does not have any explicit temporal information about the time series provided, i.e. if $X^{(j)}$ represents a population difference in the $d_p$ space, different ordering of the elements of $X^{(j)}$  will produce the same attention matrix $Z$. One way to introduce the temporal information in the self-attention layer is to explicitly add it to the input.
Then, it is necessary to create a representation of the time using a positional encoder (PE) and add it to the representation of the population difference in the $d_p$ dimensional space $X= P + \text{PE} $, with $\text{PE} \in \mathbb{R}^{T \times d_p}$ is the representation of the time with the positional encoding and $P \in \mathbb{R}^{T \times d_p}$ is the representation of the population difference that will be generated using an FFNN.  Note that the positions range from 1 to $\mathrm{T}$, where $\mathrm{T}$ is the length of the input vector.

\begin{figure}
    \centering
    \includegraphics[width=0.8\columnwidth]{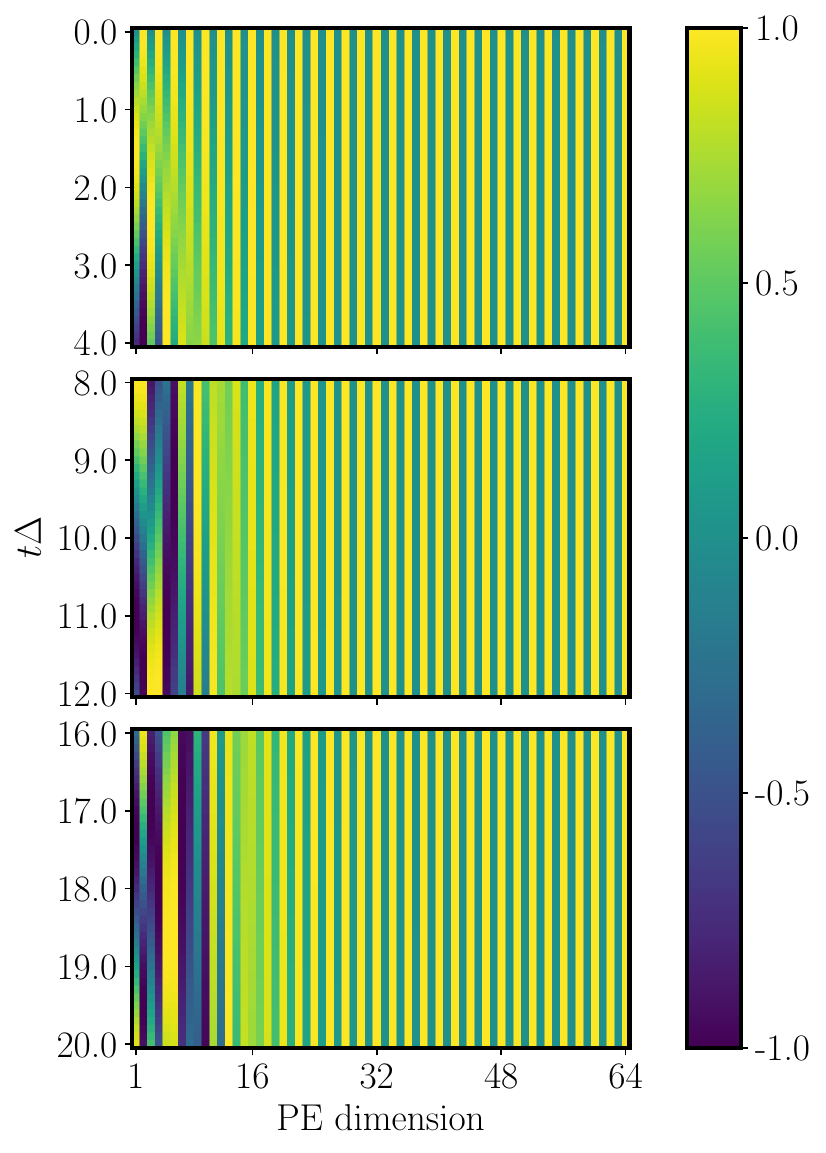}
    \caption{Positional embedding for three different time series. From top to bottom: $\textbf{t} = (0,0.1,0.2, \dots,4.0)$,  $\textbf{t} = (8.0,8.1,8.2, \dots,12.0)$, and $\textbf{t} = (16.0,16.1, \dots,20.0)$. Colors correspond to the magnitude of the embedding given by the trigonometric functions, Eq.\eqref{eq:PE}.   Each time step is projected into a $d_p=64$ dimensional space where low dimensions have the positional information and high dimensions are constant in time.}
    \label{fig:PE}
\end{figure}

The $\mathrm{PE}$ consists of trigonometric transformation based on the value of the time with different frequencies $\omega_k$ 

\begin{equation}
    \text{PE}_{j,k}= \begin{cases}\sin \left(t^{(j)} \cdot \omega_k\right) & k \text { is even } \\ \cos \left(t^{(j)} \cdot \omega_k\right) & k \text { is odd }\end{cases},
    \label{eq:PE}
\end{equation}
where the $j \in \left[ 1, \dots , T\right]$, $k \in\left[0, \ldots, d_{p}-1\right]$, $d_p$ is the dimensionality of the PE, and the frequencies are given by  
\begin{equation}
    \omega_k=\frac{1}{1000^\frac{2 k}{d_{p}}} .\label{eq:ww}
\end{equation}
Trigonometric functions are bounded  within $[-1,1]$ and  capture the periodic behavior, they span wavelengths from $2\pi (k=0)$ up to $2\pi \times 1000^2 (k= d_p)$
as in the original implementation of the transformer model.\cite{vaswani2017attention} One of the advantages of the PE based on trigonometric functions  is the notion of relative position, since a column vector PE$_{t+k} \in \mathbb{R}^{d_p}$ for a fixed time of the PE can be represented as a linear transformation PE$_{t+k}$ = MP$_t$, where $M$ is the transformation matrix.
The effect of having different frequencies $\omega_k$ is to allow the representation of the position to be unique for any projected dimension $d_p$.
This PE is non-trainable but it can be extended to the trainable PE.\cite{moreno2023positional} Furthermore, this PE is different from the original implementation  of the transformer model. In our model the PE depends on the actual value of the time $t^{(j)}$ rather than on the relative position $(j)$ in the time sequence. This choice is motivated by the Astromer architecture. Additionally, we note that in the Astromer model a factor of 1000 is used in the denominator of Eq.~\eqref{eq:ww} instead of 10000 used in the original transformer model.
The value of the PE at $k$th dimension is controlled by the  $k$th angular frequency, this makes the smaller (larger) frequencies control the  higher (smaller) dimension of the PE. In Fig. \ref{fig:PE} we show the PE for some time inputs $\mathbf{t}$ in different ranges of the evolution.  

\textit{Data sets.}
The data set used here is taken from Ref.~\citenum{ullah2023qd3set}. We describe it here for completeness. The reduced density matrix of the spin-boson Hamiltonian is calculated 
for all combinations of the following parameters: $\epsilon / \Delta=\{0,1\}$,\\ $\lambda / \Delta=\{0.1,0.2,0.3,0.4,0.5,0.6,0.7,0.8,0.9,1.0\},\; \omega_c / \Delta=\{1,2,3,4,5,6,7,8,9,10\}$, and $\beta \Delta=\{0.1,0.25,0.5,0.75,1\}$.
HEOM method implemented in QuTiP software package\cite{johansson2012qutip} was used in all calculations.  The total propagation time was $t_{\max } \Delta=20$. The integration time-step  was set to $t \Delta=0.05$. In total, 1000 HEOM calculations, 500 for symmetric $(\epsilon / \Delta=0)$ and 500 for asymmetric $(\epsilon / \Delta=1)$ spin-boson Hamiltonian were performed. Time-evolved subsystem reduced density matrices were saved every $\mathrm{d} t \Delta=0.1$. The population differences $\left\langle\hat{\sigma}_z(t)\right\rangle$ are calculated from the reduced denisty matrices and processed into shorter sequences of length $T$ by window slicing. \cite{herrera2021convolutional,ullah2021speeding,lin2021simulation}
Namely, for a time series $\mathbf{x}=\left(x^{(1)}, \ldots, x^{(L)}\right)$, where $\left\langle\hat{\sigma}_z(t^{(j)})\right\rangle$ is denoted by $x^{(j)}$ for compactness, a slice is a subset of the original time series defined as $\mathbf{s}_{i: j}=\left(x^{(i)}, \ldots, x^{(j)}\right)$ where $1 \leqslant i \leqslant j \leqslant P$. For a given time series $\mathbf{x}$ of length $L$, and the length of the slice $P$, a set of $L-P+1$ sliced time series $\left\{\mathbf{s}_{1: P}, \mathbf{s}_{2: P+1}, \ldots, \mathbf{s}_{L-P+1: L}\right\}$ was generated. This procedure was also applied to the times $\mathbf{t}$ of the original sequence. Finally, the total data set $\mathcal{D}=\left\{\left( \{ \mathbf{x}_i, \mathbf{t}_i \} , y_i\right)\right\}_{i=1}^N$ containing time series $\mathbf{x}_i$  and $\mathbf{t}_i$  with their corresponding labels $y_i$ is obtained by setting $1, \ldots, T$ elements of each slice, with $T=P-1$, to an input time-series $\mathbf{x}_i$ and the last ($P$th) element of each slice to the associated label $y_i$.

In general, the size of the window $P-1$ or, equivalently $T$, should be treated as a hyperparameter but, following our previous works, \cite{herrera2021convolutional,rodriguez22} we set $T=0.2 L$. Window slicing is applied to all 1000 HEOM reduced density matrices with different system and system-bath parameters. For each set of parameters, the initially calculated set of time-evolved $\left\langle\hat{\sigma}_z(t)\right\rangle$ with $L=t_{\max } / \mathrm{d} t=200$ generates 160 data points for the data set with $T=41$ (including $t \Delta=0$ point).

From the raw HEOM data set of 1000 trajectories, 100 randomly chosen trajectories are taken as the hold-out test set, which is used for testing and generating the results. The remaining set of 900 trajectories is transformed into 144,000 trajectories by window slicing as described above. In total 72,000 short-time $\left\langle\hat{\sigma}_z(t)\right\rangle$ trajectories for symmetric and 72,000 for asymmetric spin-boson models were generated. Each trajectory has a time length of $t \Delta+\mathrm{d} t \Delta=4.1$ where $t \Delta=4.0$ is a part of a trajectory used as input vector $\mathbf{x}_i$.
This length of an input trajectory was chosen to allow for a direct comparison to previously published ML models that use the same input.\cite{herrera2021convolutional,rodriguez22} As we showed previously, the longer the input trajectory the more accurate the prediction.\cite{herrera2021convolutional} However, generating a longer input trajectory is more computationally demanding. Therefore when choosing this parameter 
one has to balance the tradeoff between the cost of generating input data and the  prediction accuracy. The last point $t \Delta+\mathrm{d} t \Delta=4.1$ is used as a label $y_i$. 
Following previous similar works  the input data was not normalized.\cite{herrera2021convolutional,ullah2021speeding} This data set of supervised trajectories is the training set which is randomly partitioned into two subsets: a sub-training set, which contains $80 \%$ of the data and a validation set containing $20 \%$ of the data. The transformer model parameters, weights and biases, were initially fitted on the sub-training set and the validation set is used for monitoring the performance of the models (mainly to prevent overfitting). 

\begin{figure}
    \centering
    \includegraphics[width=0.8\columnwidth]{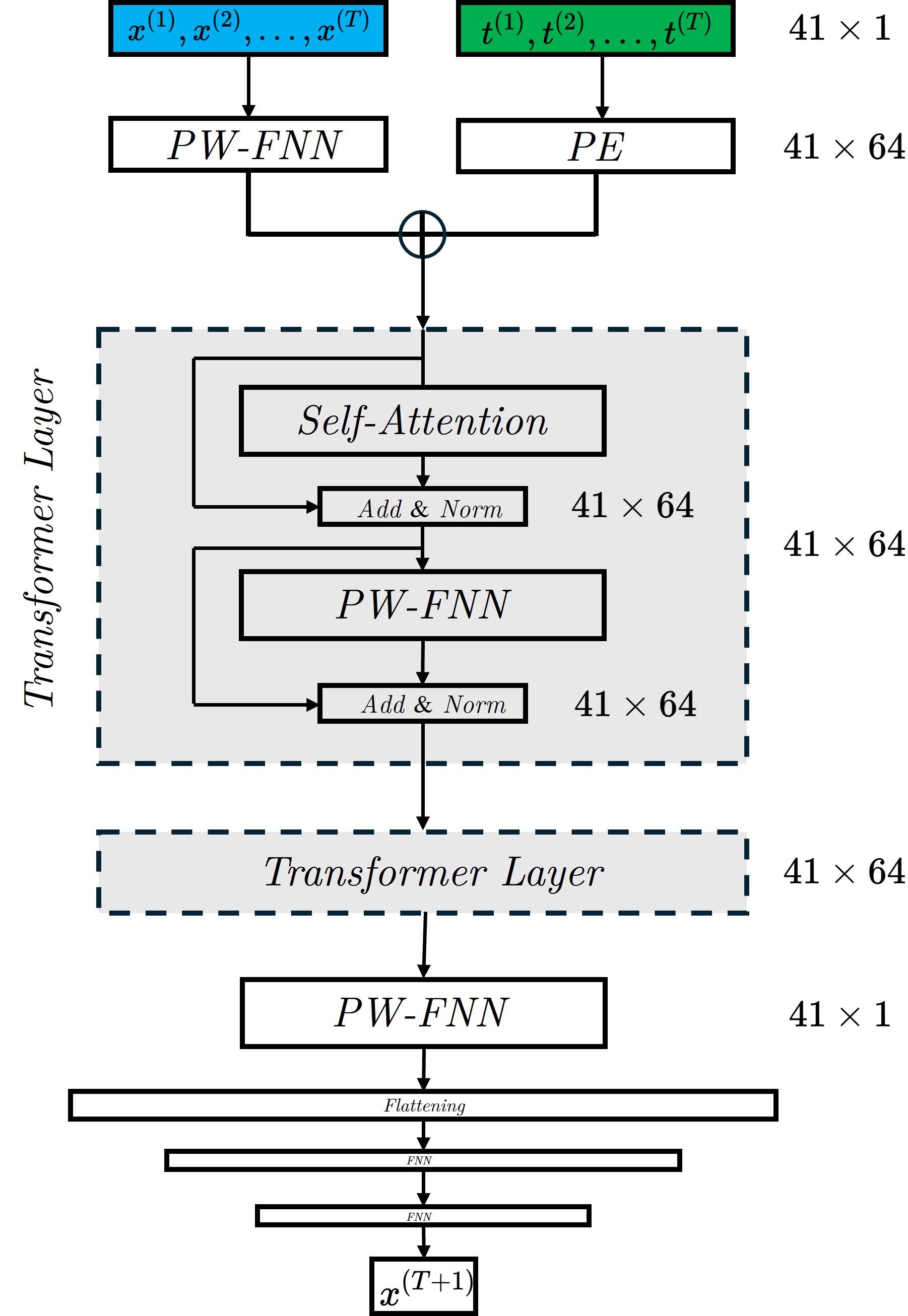}
    \caption{ML model architecture. The inputs correspond to the population difference $\mathbf{x}$ (blue) and a vector of corresponding times $\mathbf{t}$ (green). The model contains two transformer layers (dashed boxes).  The output corresponds to the population difference at the time step immediately following the last time step of the input: $x^{(T+1)} = \langle \hat{\sigma}_z(T+1) \rangle$. }
    \label{fig:Model}
\end{figure}

\textit{Transformer Model}. 
The transformer-based model employed here is inspired by the ``Astromer'' model.\cite{donoso2023astromer} 
The model architecture is shown in Fig. \ref{fig:Model}. The input is the fixed-length trajectories corresponding to the population difference  $\textbf{x}$ and the corresponding time values $\textbf{t}$. In order to use the self-attention mechanism, the inputs are embedded in objects $X \in \mathbb{R}^{T\times d_p}$ that represent the original information of the time series in a $d_p$ dimensional space. After the self-attention block the input $X$ will be transformed into $Z \in \mathbb{R}^{T\times d_p}$. The population differences  are projected into the $d_p = 1 \times 64 $ space using a linear position-wise feed-forward neural network (PW-FFNN) without hidden layers. This operation transforms the population difference into a vector of size $d_p$. The time values are projected using the PE. At each step the PE is calculated using Eq.~\eqref{eq:PE} projecting the time value into a  vector of $d_p$. Note the match between the projected dimension of the FFNN for the population difference and the PE, doing so the two inputs are encoded in the  space of the same dimensionality, and can be added as a single input for the self-attention layer. The FFNN learns how to encode the values of the population difference without interfering with the constant part of the PE. This means that the population difference will be represented in the higher dimension of $X= P +PE$ since the higher dimensions of the PE are constant as shown in Fig \ref{fig:PE}. After the addition, the inputs will be represented by the matrix $X$ of $41 \times 64$.

$X$ is followed by the two transformer layers. The transformer layer comprises a self-attention block followed by a residual connection and a normalization layer. The self-attention block  transforms the input $X$  representation into $Z \in \mathbb{R}^{41 \times \text{$\#$heads} \; d_v}$, where the number of head was set $\text{$\#$heads} =1 $  and  the $d_v=64$. After the self-attention layer a PW-FFNN with one hidden layer containing 1536 neurons and \textit{tanh} as the activation function is followed. These layers are followed by a residual connection and a normalization layer. 

After the transformer layers, in order to reduce the dimensions before  flattening, a PW-FNN is applied. After  flattening, two fully connected FFNNs are used with a number of neurons of 1024 and 1408 respectively and RELU  as the activation function. The final output of the ML model is a single neuron that gives the population difference at the next time step $\langle \hat{\sigma} (T+1)\rangle$.

The model was implemented using Keras \cite{keras} software with TensorFlow blackened.\cite{tensorflow} The model was trained using the adaptive moment estimation (Adam) algorithm \cite{diederik2014adam} with an initial learning rate of $1.0 \times 10 ^{-4}$. The batch size was set to $N_b = 128$ and the mean square error (MSE) was used as the loss function. The hyperparameters of the model: $\#$heads, $d_p$, the number transformer layers, and neurons in the fully-connected layers were optimized on the  hold-out set for the asymmetric spin-boson model using KerasTuner\cite{omalley2019kerastuner} for 50 iterations of Bayesian optimization. 500 epochs were used in the training. After the best model was chosen it was further trained for up to 4,500 epochs and the model with the lowest validation loss during the training was used. The MSE of the validation set was below $10^{-6}-10^{-7}$. This model was used to predict the population dynamics for the asymmetric spin-boson model.

To generate the population dynamics for the symmetric spin-boson model, we retrained the model described above (for the asymmetric spin-boson model) based on the data set for the symmetric spin-boson Hamiltonian. The same hyperparameters and the initial weights were taken from the already trained asymmetric model. Since the asymmetric model already learned an adequate representation of input trajectories for the asymmetric spin-boson model, the symmetric ML model was trained only for 1000 epochs and the best model was chosen again based on validation MSE. The training was performed on a single NVIDIA RTX A6000 graphics card.


Starting from the input sequence of $\mathbf{x}=\left( x^{(1)},\ldots,x^{(T)}\right)$ one uses a ML model to generate the population difference at $T+1$ (or $P$th) time step $y$. The newly predicted population difference is then combined with $T$ previous population differences to form another input vector $\mathbf{x}=\left( x^{(2)},\ldots,x^{(T)},y\to x^{(T+1)}\right)$ which is fed into the model again to generate the next prediction $y \to x^{(T+2)}$. Repeating this procedure will generate the long-time population dynamics from the initial short-time input.
It should be noted that the next time value should match the time sequence, i.e. any time point should be $t^{(n)} = (n-1)dt$, and it is necessary to update the time input correspondingly. 

\begin{figure}
    \centering
    \includegraphics[width=0.99\textwidth]{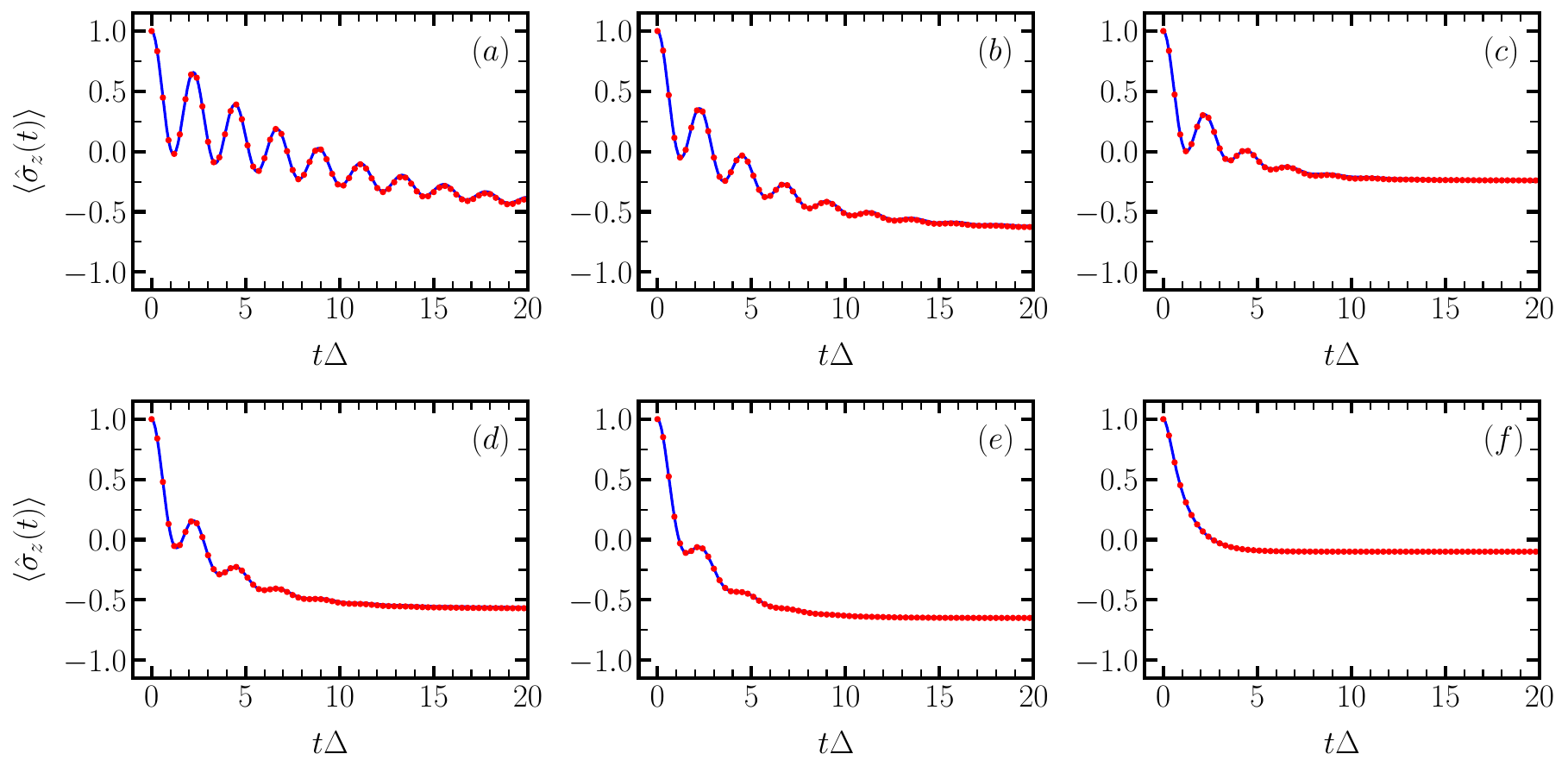}
    \caption{Expectation values $\left\langle\hat{\sigma}_z(t)\right\rangle$ for the asymmetric spin-boson model with $\epsilon=1$ as a function of time. Results predicted by the transformer-based ML model developed in this work (red circles) are compared to the numerically exact HEOM results (blue) for the following parameters: (a) $\lambda=0.1, \omega_c=6.0,
\beta=0.75$; (b)  $\lambda=0.3, \omega_c=8.0, \beta=1.0$; (c)  $\lambda=0.2, \omega_c=10.0, \beta=0.25$; (d)  $\lambda=0.4, \omega_c=8.0, \beta=0.75$; (e) $\lambda=0.8 ,  \omega_c=10.0, \beta=1.0$; (f) $\lambda=0.7, \omega_c=10.0, \beta=0.1 $. All parameters are in the units of  $\Delta$.}
    \label{fig:asym}
\end{figure}

In Fig. \ref{fig:asym} we show the predictions of the population difference of our transformer model for the asymmetric spin-boson system for six representative sets of parameters. The agreement between ML-generated dynamics and numerically exact HEOM dynamics is excellent. The average error (MAE) between the predicted and reference trajectories for the
entire hold-out set of 50 diverse reduced density matrix trajectories
 is $7.45 \times 10^{-3}$.  We stress that only a short trajectory of length $t \Delta =4 $ is used as an input and these trajectories were not present in the training data set. To put this value into perspective, the error achieved in this work with the transformer architecture is lower than the lowest error of any NN-based models trained for the same system 
 in our previous comprehensive study of ML models for long-time quantum dissipative dynamics.\cite{rodriguez22} 
 There the lowest error of $2.14 \times 10 ^{-2}$ was achieved for a convolutional gated recurrent unit (CGRU) model. In that work we also established that KRR models can be more accurate than their neural network counterparts. 
 Importantly, the performance of our transformer-based model is now at the level of KRR models yet by using neural networks we can avoid the known issues associated with KRR models such as poor scaling with the amount of the training data.
 However, it is important to note that the number of trainable parameters in the case the transformer-based model presented here is  $1,918,018$ which is four times the number of parameters of the convolutional GRU model. Increasing the number of parameters, however, does not necessarily guarantee the improvement in the performance of the model.

\begin{figure}
    \centering
    \includegraphics[width=0.99\textwidth]{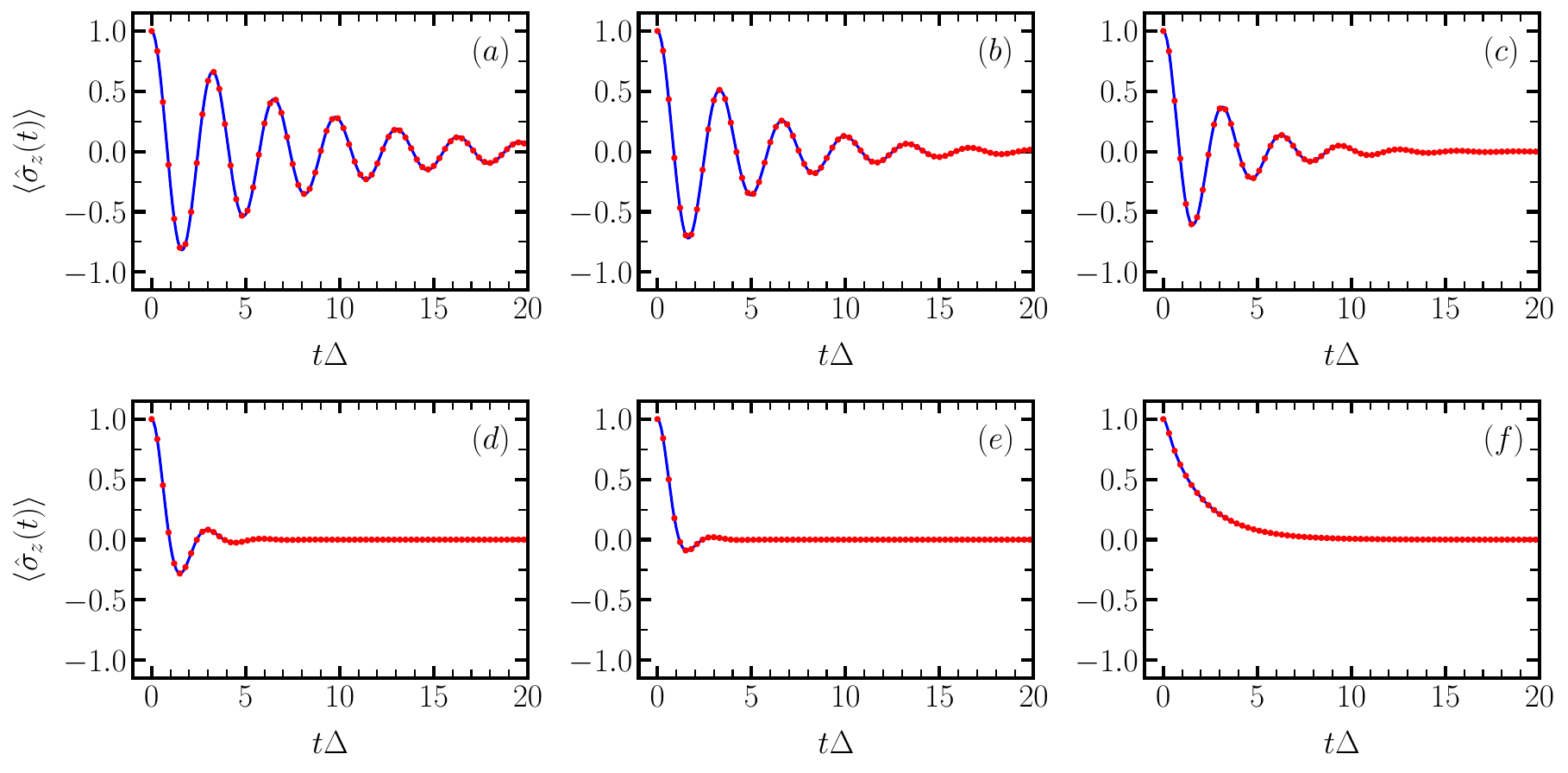}
    \caption{Expectation values $\left\langle\hat{\sigma}_z(t)\right\rangle$ for the symmetric spin-boson model as a function of time. Results predicted by the transformer-based ML model developed in this work (red circles) are compared to the numerically exact HEOM results (blue) for the following parameters: (a) $\lambda=0.2, \omega_c=8.0, \beta=1.0$;
(b) $\lambda=0.4, \omega_c=10.0, \beta=1.0$; (c) $\lambda=0.2, \omega_c=10.0, \beta=0.25$; (d) $\lambda=0.1, \omega_c=4.0, \beta=0.1$; (e) $\lambda=0.8, \omega_c=3.0, \beta=1.0$;
(f) $\lambda=1.0, \omega_c=2.0, \beta=0.1$. All parameters are in the units of $\Delta$.}
    \label{fig:sym}
\end{figure}

Fig.~\ref{fig:sym} illustrates the results for the symmetric spin-boson model. Similarly to the asymmetric spin-boson case the agreement with the HEOM is excellent. The error for the entire 50 hold-out trajectories is $4.3 \times 10^{-4}$. An even lower error in this case is expected as the dynamics of the symmetric spin-boson model is less rich compared to the asymmetric case and, thus, is easier to learn.


In summary, we showed that a transformer model trained on a set of time-discretized values of the population difference of an open quantum system is capable of predicting the future time-evolution of a given trajectory with high accuracy across the non-Makorvian and strong coupling regimes. Only a short-time trajectory generated using a numerically accurate method is required and our ML model can predict the long-time dynamics. This is the first implementation of the transformer model for the long-time dynamics of an open quantum system. It is yet another example of the ability of a ML model to predict complex physical phenomena with a lower computational cost compared to physics-based method and high accuracy.  The self-attention mechanism is a valuable tool that outperforms many popular time-forecasting models such as RNNs. The value of the self-attention mechanics is in the ability to extract long-range correlations between the time points, in contrast to RNN which have limited memory.  Additionally, the transformer layer can be executed in parallel, in contrast to the sequential of RNNs, enabling more efficient and faster training. Our transformer-based model is already accurate for the spin-boson Hamiltonian but its extensions to more complex (e.g., multi-level) systems might require using more elaborate architectures incorporating a complex decoder, such as another transformer layer that is masked for future time points or a RNN such as long short-term memory (LSTM) or GRU.

This work was supported by the U.S. Department of Energy,
Office of Science, Office of Basic Energy Sciences under
Award Number DE-SC0024511. L.E.H.R. thanks La Serena School for Data Science: Applied Tools for Astroinformatics, Biomedical Informatics, and Other Data-driven Sciences funded by NSF (AST-1637359).
A.A.K. also acknowledges NVIDIA Academic Hardware Grant Program. 

\section*{AUTHOR DECLARATIONS}
\subsection*{Conflict of Interest}
There are no conflicts of interest to disclose.

\subsection*{Author Contributions}
\textbf{Luis E. Herrera Rodríguez}: Conceptualization (equal); Methodology (lead); Investigation (lead); Writing – original draft (lead).

\textbf{Alexei A. Kananenka}: Conceptualization
(equal); Funding acquisition (lead); Methodology (equal); Writing – review \& editing (lead).

\section*{Data availability}
The source code used in this study is available at \url{https://github.com/kananenka-group/Transformer-spin-boson} under release v1.0. The data set used in this work was published elsewhere and can be accessed at 
\url{https://figshare.com/s/ed24594205ab87404238}.




\bibliography{apssamp}

\begin{thebibliography}{61}%
\makeatletter
\providecommand \@ifxundefined [1]{%
 \@ifx{#1\undefined}
}%
\providecommand \@ifnum [1]{%
 \ifnum #1\expandafter \@firstoftwo
 \else \expandafter \@secondoftwo
 \fi
}%
\providecommand \@ifx [1]{%
 \ifx #1\expandafter \@firstoftwo
 \else \expandafter \@secondoftwo
 \fi
}%
\providecommand \natexlab [1]{#1}%
\providecommand \enquote  [1]{``#1''}%
\providecommand \bibnamefont  [1]{#1}%
\providecommand \bibfnamefont [1]{#1}%
\providecommand \citenamefont [1]{#1}%
\providecommand \href@noop [0]{\@secondoftwo}%
\providecommand \href [0]{\begingroup \@sanitize@url \@href}%
\providecommand \@href[1]{\@@startlink{#1}\@@href}%
\providecommand \@@href[1]{\endgroup#1\@@endlink}%
\providecommand \@sanitize@url [0]{\catcode `\\12\catcode `\$12\catcode `\&12\catcode `\#12\catcode `\^12\catcode `\_12\catcode `\%12\relax}%
\providecommand \@@startlink[1]{}%
\providecommand \@@endlink[0]{}%
\providecommand \url  [0]{\begingroup\@sanitize@url \@url }%
\providecommand \@url [1]{\endgroup\@href {#1}{\urlprefix }}%
\providecommand \urlprefix  [0]{URL }%
\providecommand \Eprint [0]{\href }%
\providecommand \doibase [0]{http://dx.doi.org/}%
\providecommand \selectlanguage [0]{\@gobble}%
\providecommand \bibinfo  [0]{\@secondoftwo}%
\providecommand \bibfield  [0]{\@secondoftwo}%
\providecommand \translation [1]{[#1]}%
\providecommand \BibitemOpen [0]{}%
\providecommand \bibitemStop [0]{}%
\providecommand \bibitemNoStop [0]{.\EOS\space}%
\providecommand \EOS [0]{\spacefactor3000\relax}%
\providecommand \BibitemShut  [1]{\csname bibitem#1\endcsname}%
\let\auto@bib@innerbib\@empty
\bibitem [{\citenamefont {Weiss}(2012)}]{weiss2012quantum}%
  \BibitemOpen
  \bibfield  {author} {\bibinfo {author} {\bibfnamefont {U.}~\bibnamefont {Weiss}},\ }\href@noop {} {\emph {\bibinfo {title} {Quantum dissipative systems}}}\ (\bibinfo  {publisher} {World Scientific},\ \bibinfo {year} {2012})\BibitemShut {NoStop}%
\bibitem [{\citenamefont {Breuer}\ and\ \citenamefont {Petruccione}(2002)}]{breuer2002theory}%
  \BibitemOpen
  \bibfield  {author} {\bibinfo {author} {\bibfnamefont {H.-P.}\ \bibnamefont {Breuer}}\ and\ \bibinfo {author} {\bibfnamefont {F.}~\bibnamefont {Petruccione}},\ }\href@noop {} {\emph {\bibinfo {title} {The theory of open quantum systems}}}\ (\bibinfo  {publisher} {OUP Oxford},\ \bibinfo {year} {2002})\BibitemShut {NoStop}%
\bibitem [{\citenamefont {Tanimura}\ and\ \citenamefont {Kubo}(1989)}]{tanimura1989two}%
  \BibitemOpen
  \bibfield  {author} {\bibinfo {author} {\bibfnamefont {Y.}~\bibnamefont {Tanimura}}\ and\ \bibinfo {author} {\bibfnamefont {R.}~\bibnamefont {Kubo}},\ }\bibfield  {title} {\enquote {\bibinfo {title} {Two-time correlation functions of a system coupled to a heat bath with a gaussian-markoffian interaction},}\ }\href@noop {} {\bibfield  {journal} {\bibinfo  {journal} {Journal of the Physical Society of Japan}\ }\textbf {\bibinfo {volume} {58}},\ \bibinfo {pages} {1199--1206} (\bibinfo {year} {1989})}\BibitemShut {NoStop}%
\bibitem [{\citenamefont {Tanimura}(2020)}]{tanimura2020numerically}%
  \BibitemOpen
  \bibfield  {author} {\bibinfo {author} {\bibfnamefont {Y.}~\bibnamefont {Tanimura}},\ }\bibfield  {title} {\enquote {\bibinfo {title} {Numerically “exact” approach to open quantum dynamics: The hierarchical equations of motion (heom)},}\ }\href@noop {} {\bibfield  {journal} {\bibinfo  {journal} {The Journal of chemical physics}\ }\textbf {\bibinfo {volume} {153}} (\bibinfo {year} {2020})}\BibitemShut {NoStop}%
\bibitem [{\citenamefont {Wang}\ and\ \citenamefont {Thoss}(2003)}]{wang2003multilayer}%
  \BibitemOpen
  \bibfield  {author} {\bibinfo {author} {\bibfnamefont {H.}~\bibnamefont {Wang}}\ and\ \bibinfo {author} {\bibfnamefont {M.}~\bibnamefont {Thoss}},\ }\bibfield  {title} {\enquote {\bibinfo {title} {Multilayer formulation of the multiconfiguration time-dependent hartree theory},}\ }\href@noop {} {\bibfield  {journal} {\bibinfo  {journal} {The Journal of chemical physics}\ }\textbf {\bibinfo {volume} {119}},\ \bibinfo {pages} {1289--1299} (\bibinfo {year} {2003})}\BibitemShut {NoStop}%
\bibitem [{\citenamefont {Meyer}, \citenamefont {Manthe},\ and\ \citenamefont {Cederbaum}(1990)}]{meyer1990multi}%
  \BibitemOpen
  \bibfield  {author} {\bibinfo {author} {\bibfnamefont {H.-D.}\ \bibnamefont {Meyer}}, \bibinfo {author} {\bibfnamefont {U.}~\bibnamefont {Manthe}}, \ and\ \bibinfo {author} {\bibfnamefont {L.~S.}\ \bibnamefont {Cederbaum}},\ }\bibfield  {title} {\enquote {\bibinfo {title} {The multi-configurational time-dependent hartree approach},}\ }\href@noop {} {\bibfield  {journal} {\bibinfo  {journal} {Chemical Physics Letters}\ }\textbf {\bibinfo {volume} {165}},\ \bibinfo {pages} {73--78} (\bibinfo {year} {1990})}\BibitemShut {NoStop}%
\bibitem [{\citenamefont {Kundu}\ and\ \citenamefont {Makri}(2023)}]{kundu2023pathsum}%
  \BibitemOpen
  \bibfield  {author} {\bibinfo {author} {\bibfnamefont {S.}~\bibnamefont {Kundu}}\ and\ \bibinfo {author} {\bibfnamefont {N.}~\bibnamefont {Makri}},\ }\bibfield  {title} {\enquote {\bibinfo {title} {Pathsum: A c++ and fortran suite of fully quantum mechanical real-time path integral methods for (multi-) system+ bath dynamics},}\ }\href@noop {} {\bibfield  {journal} {\bibinfo  {journal} {The Journal of Chemical Physics}\ }\textbf {\bibinfo {volume} {158}} (\bibinfo {year} {2023})}\BibitemShut {NoStop}%
\bibitem [{\citenamefont {Makarov}\ and\ \citenamefont {Makri}(1994)}]{makarov1994path}%
  \BibitemOpen
  \bibfield  {author} {\bibinfo {author} {\bibfnamefont {D.~E.}\ \bibnamefont {Makarov}}\ and\ \bibinfo {author} {\bibfnamefont {N.}~\bibnamefont {Makri}},\ }\bibfield  {title} {\enquote {\bibinfo {title} {Path integrals for dissipative systems by tensor multiplication. condensed phase quantum dynamics for arbitrarily long time},}\ }\href@noop {} {\bibfield  {journal} {\bibinfo  {journal} {Chemical physics letters}\ }\textbf {\bibinfo {volume} {221}},\ \bibinfo {pages} {482--491} (\bibinfo {year} {1994})}\BibitemShut {NoStop}%
\bibitem [{\citenamefont {Luo}\ \emph {et~al.}(2010)\citenamefont {Luo}, \citenamefont {Ye}, \citenamefont {Guan},\ and\ \citenamefont {Zhao}}]{luo2010validity}%
  \BibitemOpen
  \bibfield  {author} {\bibinfo {author} {\bibfnamefont {B.}~\bibnamefont {Luo}}, \bibinfo {author} {\bibfnamefont {J.}~\bibnamefont {Ye}}, \bibinfo {author} {\bibfnamefont {C.}~\bibnamefont {Guan}}, \ and\ \bibinfo {author} {\bibfnamefont {Y.}~\bibnamefont {Zhao}},\ }\bibfield  {title} {\enquote {\bibinfo {title} {Validity of time-dependent trial states for the holstein polaron},}\ }\href@noop {} {\bibfield  {journal} {\bibinfo  {journal} {Physical Chemistry Chemical Physics}\ }\textbf {\bibinfo {volume} {12}},\ \bibinfo {pages} {15073--15084} (\bibinfo {year} {2010})}\BibitemShut {NoStop}%
\bibitem [{\citenamefont {Ren}, \citenamefont {Shuai},\ and\ \citenamefont {Kin-Lic~Chan}(2018)}]{ren2018time}%
  \BibitemOpen
  \bibfield  {author} {\bibinfo {author} {\bibfnamefont {J.}~\bibnamefont {Ren}}, \bibinfo {author} {\bibfnamefont {Z.}~\bibnamefont {Shuai}}, \ and\ \bibinfo {author} {\bibfnamefont {G.}~\bibnamefont {Kin-Lic~Chan}},\ }\bibfield  {title} {\enquote {\bibinfo {title} {Time-dependent density matrix renormalization group algorithms for nearly exact absorption and fluorescence spectra of molecular aggregates at both zero and finite temperature},}\ }\href@noop {} {\bibfield  {journal} {\bibinfo  {journal} {Journal of Chemical Theory and Computation}\ }\textbf {\bibinfo {volume} {14}},\ \bibinfo {pages} {5027--5039} (\bibinfo {year} {2018})}\BibitemShut {NoStop}%
\bibitem [{\citenamefont {Greene}\ and\ \citenamefont {Batista}(2017)}]{greene2017tensor}%
  \BibitemOpen
  \bibfield  {author} {\bibinfo {author} {\bibfnamefont {S.~M.}\ \bibnamefont {Greene}}\ and\ \bibinfo {author} {\bibfnamefont {V.~S.}\ \bibnamefont {Batista}},\ }\bibfield  {title} {\enquote {\bibinfo {title} {Tensor-train split-operator fourier transform (tt-soft) method: Multidimensional nonadiabatic quantum dynamics},}\ }\href@noop {} {\bibfield  {journal} {\bibinfo  {journal} {Journal of chemical theory and computation}\ }\textbf {\bibinfo {volume} {13}},\ \bibinfo {pages} {4034--4042} (\bibinfo {year} {2017})}\BibitemShut {NoStop}%
\bibitem [{\citenamefont {Yan}\ and\ \citenamefont {Shao}(2016)}]{yan2016stochastic}%
  \BibitemOpen
  \bibfield  {author} {\bibinfo {author} {\bibfnamefont {Y.-A.}\ \bibnamefont {Yan}}\ and\ \bibinfo {author} {\bibfnamefont {J.}~\bibnamefont {Shao}},\ }\bibfield  {title} {\enquote {\bibinfo {title} {Stochastic description of quantum brownian dynamics},}\ }\href@noop {} {\bibfield  {journal} {\bibinfo  {journal} {Frontiers of Physics}\ }\textbf {\bibinfo {volume} {11}},\ \bibinfo {pages} {1--24} (\bibinfo {year} {2016})}\BibitemShut {NoStop}%
\bibitem [{\citenamefont {Hsieh}\ and\ \citenamefont {Cao}(2018)}]{hsieh2018unified}%
  \BibitemOpen
  \bibfield  {author} {\bibinfo {author} {\bibfnamefont {C.-Y.}\ \bibnamefont {Hsieh}}\ and\ \bibinfo {author} {\bibfnamefont {J.}~\bibnamefont {Cao}},\ }\bibfield  {title} {\enquote {\bibinfo {title} {A unified stochastic formulation of dissipative quantum dynamics. i. generalized hierarchical equations},}\ }\href@noop {} {\bibfield  {journal} {\bibinfo  {journal} {The Journal of Chemical Physics}\ }\textbf {\bibinfo {volume} {148}} (\bibinfo {year} {2018})}\BibitemShut {NoStop}%
\bibitem [{\citenamefont {Han}\ \emph {et~al.}(2020)\citenamefont {Han}, \citenamefont {Ullah}, \citenamefont {Yan}, \citenamefont {Zheng}, \citenamefont {Yan},\ and\ \citenamefont {Chernyak}}]{han2020stochastic}%
  \BibitemOpen
  \bibfield  {author} {\bibinfo {author} {\bibfnamefont {L.}~\bibnamefont {Han}}, \bibinfo {author} {\bibfnamefont {A.}~\bibnamefont {Ullah}}, \bibinfo {author} {\bibfnamefont {Y.-A.}\ \bibnamefont {Yan}}, \bibinfo {author} {\bibfnamefont {X.}~\bibnamefont {Zheng}}, \bibinfo {author} {\bibfnamefont {Y.}~\bibnamefont {Yan}}, \ and\ \bibinfo {author} {\bibfnamefont {V.}~\bibnamefont {Chernyak}},\ }\bibfield  {title} {\enquote {\bibinfo {title} {Stochastic equation of motion approach to fermionic dissipative dynamics. i. formalism},}\ }\href@noop {} {\bibfield  {journal} {\bibinfo  {journal} {The Journal of Chemical Physics}\ }\textbf {\bibinfo {volume} {152}} (\bibinfo {year} {2020})}\BibitemShut {NoStop}%
\bibitem [{\citenamefont {Ullah}\ \emph {et~al.}(2020)\citenamefont {Ullah}, \citenamefont {Han}, \citenamefont {Yan}, \citenamefont {Zheng}, \citenamefont {Yan},\ and\ \citenamefont {Chernyak}}]{ullah2020stochastic}%
  \BibitemOpen
  \bibfield  {author} {\bibinfo {author} {\bibfnamefont {A.}~\bibnamefont {Ullah}}, \bibinfo {author} {\bibfnamefont {L.}~\bibnamefont {Han}}, \bibinfo {author} {\bibfnamefont {Y.-A.}\ \bibnamefont {Yan}}, \bibinfo {author} {\bibfnamefont {X.}~\bibnamefont {Zheng}}, \bibinfo {author} {\bibfnamefont {Y.}~\bibnamefont {Yan}}, \ and\ \bibinfo {author} {\bibfnamefont {V.}~\bibnamefont {Chernyak}},\ }\bibfield  {title} {\enquote {\bibinfo {title} {Stochastic equation of motion approach to fermionic dissipative dynamics. ii. numerical implementation},}\ }\href@noop {} {\bibfield  {journal} {\bibinfo  {journal} {The Journal of Chemical Physics}\ }\textbf {\bibinfo {volume} {152}} (\bibinfo {year} {2020})}\BibitemShut {NoStop}%
\bibitem [{\citenamefont {Brian}\ and\ \citenamefont {Sun}(2021)}]{brian2021generalized}%
  \BibitemOpen
  \bibfield  {author} {\bibinfo {author} {\bibfnamefont {D.}~\bibnamefont {Brian}}\ and\ \bibinfo {author} {\bibfnamefont {X.}~\bibnamefont {Sun}},\ }\bibfield  {title} {\enquote {\bibinfo {title} {Generalized quantum master equation: A tutorial review and recent advances},}\ }\href@noop {} {\bibfield  {journal} {\bibinfo  {journal} {Chinese Journal of Chemical Physics}\ }\textbf {\bibinfo {volume} {34}},\ \bibinfo {pages} {497--524} (\bibinfo {year} {2021})}\BibitemShut {NoStop}%
\bibitem [{\citenamefont {Nakajima}(1958)}]{nakajima1958quantum}%
  \BibitemOpen
  \bibfield  {author} {\bibinfo {author} {\bibfnamefont {S.}~\bibnamefont {Nakajima}},\ }\bibfield  {title} {\enquote {\bibinfo {title} {On quantum theory of transport phenomena: Steady diffusion},}\ }\href@noop {} {\bibfield  {journal} {\bibinfo  {journal} {Progress of Theoretical Physics}\ }\textbf {\bibinfo {volume} {20}},\ \bibinfo {pages} {948--959} (\bibinfo {year} {1958})}\BibitemShut {NoStop}%
\bibitem [{\citenamefont {Zwanzig}(1960)}]{zwanzig1960ensemble}%
  \BibitemOpen
  \bibfield  {author} {\bibinfo {author} {\bibfnamefont {R.}~\bibnamefont {Zwanzig}},\ }\bibfield  {title} {\enquote {\bibinfo {title} {Ensemble method in the theory of irreversibility},}\ }\href@noop {} {\bibfield  {journal} {\bibinfo  {journal} {The Journal of Chemical Physics}\ }\textbf {\bibinfo {volume} {33}},\ \bibinfo {pages} {1338--1341} (\bibinfo {year} {1960})}\BibitemShut {NoStop}%
\bibitem [{\citenamefont {Shi}\ and\ \citenamefont {Geva}(2003)}]{shi2003new}%
  \BibitemOpen
  \bibfield  {author} {\bibinfo {author} {\bibfnamefont {Q.}~\bibnamefont {Shi}}\ and\ \bibinfo {author} {\bibfnamefont {E.}~\bibnamefont {Geva}},\ }\bibfield  {title} {\enquote {\bibinfo {title} {A new approach to calculating the memory kernel of the generalized quantum master equation for an arbitrary system--bath coupling},}\ }\href@noop {} {\bibfield  {journal} {\bibinfo  {journal} {The Journal of chemical physics}\ }\textbf {\bibinfo {volume} {119}},\ \bibinfo {pages} {12063--12076} (\bibinfo {year} {2003})}\BibitemShut {NoStop}%
\bibitem [{\citenamefont {Mulvihill}\ and\ \citenamefont {Geva}(2021)}]{mulvihill2021road}%
  \BibitemOpen
  \bibfield  {author} {\bibinfo {author} {\bibfnamefont {E.}~\bibnamefont {Mulvihill}}\ and\ \bibinfo {author} {\bibfnamefont {E.}~\bibnamefont {Geva}},\ }\bibfield  {title} {\enquote {\bibinfo {title} {A road map to various pathways for calculating the memory kernel of the generalized quantum master equation},}\ }\href@noop {} {\bibfield  {journal} {\bibinfo  {journal} {The Journal of Physical Chemistry B}\ }\textbf {\bibinfo {volume} {125}},\ \bibinfo {pages} {9834--9852} (\bibinfo {year} {2021})}\BibitemShut {NoStop}%
\bibitem [{\citenamefont {Kelly}\ and\ \citenamefont {Markland}(2013)}]{kelly2013efficient}%
  \BibitemOpen
  \bibfield  {author} {\bibinfo {author} {\bibfnamefont {A.}~\bibnamefont {Kelly}}\ and\ \bibinfo {author} {\bibfnamefont {T.~E.}\ \bibnamefont {Markland}},\ }\bibfield  {title} {\enquote {\bibinfo {title} {Efficient and accurate surface hopping for long time nonadiabatic quantum dynamics},}\ }\href@noop {} {\bibfield  {journal} {\bibinfo  {journal} {The Journal of chemical physics}\ }\textbf {\bibinfo {volume} {139}} (\bibinfo {year} {2013})}\BibitemShut {NoStop}%
\bibitem [{\citenamefont {Cerrillo}\ and\ \citenamefont {Cao}(2014)}]{cerrillo2014non}%
  \BibitemOpen
  \bibfield  {author} {\bibinfo {author} {\bibfnamefont {J.}~\bibnamefont {Cerrillo}}\ and\ \bibinfo {author} {\bibfnamefont {J.}~\bibnamefont {Cao}},\ }\bibfield  {title} {\enquote {\bibinfo {title} {Non-markovian dynamical maps: numerical processing of open quantum trajectories},}\ }\href@noop {} {\bibfield  {journal} {\bibinfo  {journal} {Physical review letters}\ }\textbf {\bibinfo {volume} {112}},\ \bibinfo {pages} {110401} (\bibinfo {year} {2014})}\BibitemShut {NoStop}%
\bibitem [{\citenamefont {Kananenka}\ \emph {et~al.}(2016)\citenamefont {Kananenka}, \citenamefont {Hsieh}, \citenamefont {Cao},\ and\ \citenamefont {Geva}}]{kananenka2016accurate}%
  \BibitemOpen
  \bibfield  {author} {\bibinfo {author} {\bibfnamefont {A.~A.}\ \bibnamefont {Kananenka}}, \bibinfo {author} {\bibfnamefont {C.-Y.}\ \bibnamefont {Hsieh}}, \bibinfo {author} {\bibfnamefont {J.}~\bibnamefont {Cao}}, \ and\ \bibinfo {author} {\bibfnamefont {E.}~\bibnamefont {Geva}},\ }\bibfield  {title} {\enquote {\bibinfo {title} {Accurate long-time mixed quantum-classical liouville dynamics via the transfer tensor method},}\ }\href@noop {} {\bibfield  {journal} {\bibinfo  {journal} {The journal of physical chemistry letters}\ }\textbf {\bibinfo {volume} {7}},\ \bibinfo {pages} {4809--4814} (\bibinfo {year} {2016})}\BibitemShut {NoStop}%
\bibitem [{\citenamefont {Buser}\ \emph {et~al.}(2017)\citenamefont {Buser}, \citenamefont {Cerrillo}, \citenamefont {Schaller},\ and\ \citenamefont {Cao}}]{buser2017initial}%
  \BibitemOpen
  \bibfield  {author} {\bibinfo {author} {\bibfnamefont {M.}~\bibnamefont {Buser}}, \bibinfo {author} {\bibfnamefont {J.}~\bibnamefont {Cerrillo}}, \bibinfo {author} {\bibfnamefont {G.}~\bibnamefont {Schaller}}, \ and\ \bibinfo {author} {\bibfnamefont {J.}~\bibnamefont {Cao}},\ }\bibfield  {title} {\enquote {\bibinfo {title} {Initial system-environment correlations via the transfer-tensor method},}\ }\href@noop {} {\bibfield  {journal} {\bibinfo  {journal} {Physical Review A}\ }\textbf {\bibinfo {volume} {96}},\ \bibinfo {pages} {062122} (\bibinfo {year} {2017})}\BibitemShut {NoStop}%
\bibitem [{\citenamefont {Gelzinis}, \citenamefont {Rybakovas},\ and\ \citenamefont {Valkunas}(2017)}]{gelzinis2017applicability}%
  \BibitemOpen
  \bibfield  {author} {\bibinfo {author} {\bibfnamefont {A.}~\bibnamefont {Gelzinis}}, \bibinfo {author} {\bibfnamefont {E.}~\bibnamefont {Rybakovas}}, \ and\ \bibinfo {author} {\bibfnamefont {L.}~\bibnamefont {Valkunas}},\ }\bibfield  {title} {\enquote {\bibinfo {title} {Applicability of transfer tensor method for open quantum system dynamics},}\ }\href@noop {} {\bibfield  {journal} {\bibinfo  {journal} {The Journal of chemical physics}\ }\textbf {\bibinfo {volume} {147}} (\bibinfo {year} {2017})}\BibitemShut {NoStop}%
\bibitem [{\citenamefont {Chen}\ \emph {et~al.}(2020)\citenamefont {Chen}, \citenamefont {Ma}, \citenamefont {Zheng}, \citenamefont {Allcock}, \citenamefont {Zhang},\ and\ \citenamefont {Hsieh}}]{chen2020non}%
  \BibitemOpen
  \bibfield  {author} {\bibinfo {author} {\bibfnamefont {Y.-Q.}\ \bibnamefont {Chen}}, \bibinfo {author} {\bibfnamefont {K.-L.}\ \bibnamefont {Ma}}, \bibinfo {author} {\bibfnamefont {Y.-C.}\ \bibnamefont {Zheng}}, \bibinfo {author} {\bibfnamefont {J.}~\bibnamefont {Allcock}}, \bibinfo {author} {\bibfnamefont {S.}~\bibnamefont {Zhang}}, \ and\ \bibinfo {author} {\bibfnamefont {C.-Y.}\ \bibnamefont {Hsieh}},\ }\bibfield  {title} {\enquote {\bibinfo {title} {Non-markovian noise characterization with the transfer tensor method},}\ }\href@noop {} {\bibfield  {journal} {\bibinfo  {journal} {Physical Review Applied}\ }\textbf {\bibinfo {volume} {13}},\ \bibinfo {pages} {034045} (\bibinfo {year} {2020})}\BibitemShut {NoStop}%
\bibitem [{\citenamefont {Herrera~Rodr{\'i}guez}\ and\ \citenamefont {Kananenka}(2021)}]{herrera2021convolutional}%
  \BibitemOpen
  \bibfield  {author} {\bibinfo {author} {\bibfnamefont {L.~E.}\ \bibnamefont {Herrera~Rodr{\'i}guez}}\ and\ \bibinfo {author} {\bibfnamefont {A.~A.}\ \bibnamefont {Kananenka}},\ }\bibfield  {title} {\enquote {\bibinfo {title} {Convolutional neural networks for long time dissipative quantum dynamics},}\ }\href@noop {} {\bibfield  {journal} {\bibinfo  {journal} {The Journal of Physical Chemistry Letters}\ }\textbf {\bibinfo {volume} {12}},\ \bibinfo {pages} {2476--2483} (\bibinfo {year} {2021})}\BibitemShut {NoStop}%
\bibitem [{\citenamefont {Ullah}\ and\ \citenamefont {Dral}(2021)}]{ullah2021speeding}%
  \BibitemOpen
  \bibfield  {author} {\bibinfo {author} {\bibfnamefont {A.}~\bibnamefont {Ullah}}\ and\ \bibinfo {author} {\bibfnamefont {P.~O.}\ \bibnamefont {Dral}},\ }\bibfield  {title} {\enquote {\bibinfo {title} {Speeding up quantum dissipative dynamics of open systems with kernel methods},}\ }\href@noop {} {\bibfield  {journal} {\bibinfo  {journal} {New Journal of Physics}\ }\textbf {\bibinfo {volume} {23}},\ \bibinfo {pages} {113019} (\bibinfo {year} {2021})}\BibitemShut {NoStop}%
\bibitem [{\citenamefont {Ullah}\ and\ \citenamefont {Dral}(2022{\natexlab{a}})}]{ullah2022predicting}%
  \BibitemOpen
  \bibfield  {author} {\bibinfo {author} {\bibfnamefont {A.}~\bibnamefont {Ullah}}\ and\ \bibinfo {author} {\bibfnamefont {P.~O.}\ \bibnamefont {Dral}},\ }\bibfield  {title} {\enquote {\bibinfo {title} {Predicting the future of excitation energy transfer in light-harvesting complex with artificial intelligence-based quantum dynamics},}\ }\href@noop {} {\bibfield  {journal} {\bibinfo  {journal} {Nature Communications}\ }\textbf {\bibinfo {volume} {13}},\ \bibinfo {pages} {1930} (\bibinfo {year} {2022}{\natexlab{a}})}\BibitemShut {NoStop}%
\bibitem [{\citenamefont {Ullah}\ and\ \citenamefont {Dral}(2022{\natexlab{b}})}]{ullah2022one}%
  \BibitemOpen
  \bibfield  {author} {\bibinfo {author} {\bibfnamefont {A.}~\bibnamefont {Ullah}}\ and\ \bibinfo {author} {\bibfnamefont {P.~O.}\ \bibnamefont {Dral}},\ }\bibfield  {title} {\enquote {\bibinfo {title} {One-shot trajectory learning of open quantum systems dynamics},}\ }\href@noop {} {\bibfield  {journal} {\bibinfo  {journal} {The Journal of Physical Chemistry Letters}\ }\textbf {\bibinfo {volume} {13}},\ \bibinfo {pages} {6037--6041} (\bibinfo {year} {2022}{\natexlab{b}})}\BibitemShut {NoStop}%
\bibitem [{\citenamefont {Ullah}\ \emph {et~al.}(2023)\citenamefont {Ullah}, \citenamefont {Herrera~Rodr{\'\i}guez}, \citenamefont {Dral},\ and\ \citenamefont {Kananenka}}]{ullah2023qd3set}%
  \BibitemOpen
  \bibfield  {author} {\bibinfo {author} {\bibfnamefont {A.}~\bibnamefont {Ullah}}, \bibinfo {author} {\bibfnamefont {L.~E.}\ \bibnamefont {Herrera~Rodr{\'\i}guez}}, \bibinfo {author} {\bibfnamefont {P.~O.}\ \bibnamefont {Dral}}, \ and\ \bibinfo {author} {\bibfnamefont {A.~A.}\ \bibnamefont {Kananenka}},\ }\bibfield  {title} {\enquote {\bibinfo {title} {Qd3set-1: a database with quantum dissipative dynamics datasets},}\ }\href@noop {} {\bibfield  {journal} {\bibinfo  {journal} {Frontiers in Physics}\ }\textbf {\bibinfo {volume} {11}},\ \bibinfo {pages} {1223973} (\bibinfo {year} {2023})}\BibitemShut {NoStop}%
\bibitem [{\citenamefont {Ullah}\ and\ \citenamefont {Dral}(2024)}]{ullah2024mlqd}%
  \BibitemOpen
  \bibfield  {author} {\bibinfo {author} {\bibfnamefont {A.}~\bibnamefont {Ullah}}\ and\ \bibinfo {author} {\bibfnamefont {P.~O.}\ \bibnamefont {Dral}},\ }\bibfield  {title} {\enquote {\bibinfo {title} {Mlqd: A package for machine learning-based quantum dissipative dynamics},}\ }\href@noop {} {\bibfield  {journal} {\bibinfo  {journal} {Computer Physics Communications}\ }\textbf {\bibinfo {volume} {294}},\ \bibinfo {pages} {108940} (\bibinfo {year} {2024})}\BibitemShut {NoStop}%
\bibitem [{\citenamefont {Lin}\ \emph {et~al.}(2021)\citenamefont {Lin}, \citenamefont {Peng}, \citenamefont {Gu},\ and\ \citenamefont {Lan}}]{lin2021simulation}%
  \BibitemOpen
  \bibfield  {author} {\bibinfo {author} {\bibfnamefont {K.}~\bibnamefont {Lin}}, \bibinfo {author} {\bibfnamefont {J.}~\bibnamefont {Peng}}, \bibinfo {author} {\bibfnamefont {F.~L.}\ \bibnamefont {Gu}}, \ and\ \bibinfo {author} {\bibfnamefont {Z.}~\bibnamefont {Lan}},\ }\bibfield  {title} {\enquote {\bibinfo {title} {Simulation of open quantum dynamics with bootstrap-based long short-term memory recurrent neural network},}\ }\href@noop {} {\bibfield  {journal} {\bibinfo  {journal} {The Journal of Physical Chemistry Letters}\ }\textbf {\bibinfo {volume} {12}},\ \bibinfo {pages} {10225--10234} (\bibinfo {year} {2021})}\BibitemShut {NoStop}%
\bibitem [{\citenamefont {Wu}\ \emph {et~al.}(2021)\citenamefont {Wu}, \citenamefont {Hu}, \citenamefont {Li},\ and\ \citenamefont {Sun}}]{wu21}%
  \BibitemOpen
  \bibfield  {author} {\bibinfo {author} {\bibfnamefont {D.}~\bibnamefont {Wu}}, \bibinfo {author} {\bibfnamefont {Z.}~\bibnamefont {Hu}}, \bibinfo {author} {\bibfnamefont {J.}~\bibnamefont {Li}}, \ and\ \bibinfo {author} {\bibfnamefont {X.}~\bibnamefont {Sun}},\ }\bibfield  {title} {\enquote {\bibinfo {title} {{Forecasting nonadiabatic dynamics using hybrid convolutional neural network/long short-term memory network}},}\ }\href {\doibase 10.1063/5.0073689} {\bibfield  {journal} {\bibinfo  {journal} {The Journal of Chemical Physics}\ }\textbf {\bibinfo {volume} {155}},\ \bibinfo {pages} {224104} (\bibinfo {year} {2021})},\ \Eprint {http://arxiv.org/abs/https://pubs.aip.org/aip/jcp/article-pdf/doi/10.1063/5.0073689/15963310/224104\_1\_online.pdf} {https://pubs.aip.org/aip/jcp/article-pdf/doi/10.1063/5.0073689/15963310/224104\_1\_online.pdf} \BibitemShut {NoStop}%
\bibitem [{\citenamefont {Akimov}(2021)}]{akimov21}%
  \BibitemOpen
  \bibfield  {author} {\bibinfo {author} {\bibfnamefont {A.~V.}\ \bibnamefont {Akimov}},\ }\bibfield  {title} {\enquote {\bibinfo {title} {Extending the time scales of nonadiabatic molecular dynamics via machine learning in the time domain},}\ }\href {\doibase 10.1021/acs.jpclett.1c03823} {\bibfield  {journal} {\bibinfo  {journal} {The Journal of Physical Chemistry Letters}\ }\textbf {\bibinfo {volume} {12}},\ \bibinfo {pages} {12119--12128} (\bibinfo {year} {2021})},\ \bibinfo {note} {pMID: 34913701},\ \Eprint {http://arxiv.org/abs/https://doi.org/10.1021/acs.jpclett.1c03823} {https://doi.org/10.1021/acs.jpclett.1c03823} \BibitemShut {NoStop}%
\bibitem [{\citenamefont {Zhang}\ \emph {et~al.}(2024)\citenamefont {Zhang}, \citenamefont {Pios}, \citenamefont {Martyka}, \citenamefont {Ge}, \citenamefont {Hou}, \citenamefont {Chen}, \citenamefont {Chen}, \citenamefont {Jankowska}, \citenamefont {Barbatti},\ and\ \citenamefont {Dral}}]{zhang24}%
  \BibitemOpen
  \bibfield  {author} {\bibinfo {author} {\bibfnamefont {L.}~\bibnamefont {Zhang}}, \bibinfo {author} {\bibfnamefont {S.~V.}\ \bibnamefont {Pios}}, \bibinfo {author} {\bibfnamefont {M.}~\bibnamefont {Martyka}}, \bibinfo {author} {\bibfnamefont {F.}~\bibnamefont {Ge}}, \bibinfo {author} {\bibfnamefont {Y.-F.}\ \bibnamefont {Hou}}, \bibinfo {author} {\bibfnamefont {Y.}~\bibnamefont {Chen}}, \bibinfo {author} {\bibfnamefont {L.}~\bibnamefont {Chen}}, \bibinfo {author} {\bibfnamefont {J.}~\bibnamefont {Jankowska}}, \bibinfo {author} {\bibfnamefont {M.}~\bibnamefont {Barbatti}}, \ and\ \bibinfo {author} {\bibfnamefont {P.~O.}\ \bibnamefont {Dral}},\ }\bibfield  {title} {\enquote {\bibinfo {title} {Mlatom software ecosystem for surface hopping dynamics in python with quantum mechanical and machine learning methods},}\ }\href {\doibase 10.1021/acs.jctc.4c00468} {\bibfield  {journal} {\bibinfo  {journal} {Journal of Chemical Theory and Computation}\ }\textbf {\bibinfo {volume} {20}},\ \bibinfo {pages} {5043--5057}
  (\bibinfo {year} {2024})},\ \bibinfo {note} {pMID: 38836623},\ \Eprint {http://arxiv.org/abs/https://doi.org/10.1021/acs.jctc.4c00468} {https://doi.org/10.1021/acs.jctc.4c00468} \BibitemShut {NoStop}%
\bibitem [{\citenamefont {Lin}\ \emph {et~al.}(2022)\citenamefont {Lin}, \citenamefont {Peng}, \citenamefont {Xu}, \citenamefont {Gu},\ and\ \citenamefont {Lan}}]{lin22}%
  \BibitemOpen
  \bibfield  {author} {\bibinfo {author} {\bibfnamefont {K.}~\bibnamefont {Lin}}, \bibinfo {author} {\bibfnamefont {J.}~\bibnamefont {Peng}}, \bibinfo {author} {\bibfnamefont {C.}~\bibnamefont {Xu}}, \bibinfo {author} {\bibfnamefont {F.~L.}\ \bibnamefont {Gu}}, \ and\ \bibinfo {author} {\bibfnamefont {Z.}~\bibnamefont {Lan}},\ }\bibfield  {title} {\enquote {\bibinfo {title} {Automatic evolution of machine-learning-based quantum dynamics with uncertainty analysis},}\ }\href {\doibase 10.1021/acs.jctc.2c00702} {\bibfield  {journal} {\bibinfo  {journal} {Journal of Chemical Theory and Computation}\ }\textbf {\bibinfo {volume} {18}},\ \bibinfo {pages} {5837--5855} (\bibinfo {year} {2022})},\ \bibinfo {note} {pMID: 36184823},\ \Eprint {http://arxiv.org/abs/https://doi.org/10.1021/acs.jctc.2c00702} {https://doi.org/10.1021/acs.jctc.2c00702} \BibitemShut {NoStop}%
\bibitem [{\citenamefont {Luo}\ \emph {et~al.}(2022)\citenamefont {Luo}, \citenamefont {Chen}, \citenamefont {Carrasquilla},\ and\ \citenamefont {Clark}}]{luo2009autoregressive}%
  \BibitemOpen
  \bibfield  {author} {\bibinfo {author} {\bibfnamefont {D.}~\bibnamefont {Luo}}, \bibinfo {author} {\bibfnamefont {Z.}~\bibnamefont {Chen}}, \bibinfo {author} {\bibfnamefont {J.}~\bibnamefont {Carrasquilla}}, \ and\ \bibinfo {author} {\bibfnamefont {B.~K.}\ \bibnamefont {Clark}},\ }\bibfield  {title} {\enquote {\bibinfo {title} {Autoregressive neural network for simulating open quantum systems via a probabilistic formulation},}\ }\href {\doibase 10.1103/PhysRevLett.128.090501} {\bibfield  {journal} {\bibinfo  {journal} {Phys. Rev. Lett.}\ }\textbf {\bibinfo {volume} {128}},\ \bibinfo {pages} {090501} (\bibinfo {year} {2022})}\BibitemShut {NoStop}%
\bibitem [{\citenamefont {Rodríguez}\ \emph {et~al.}(2022)\citenamefont {Rodríguez}, \citenamefont {Ullah}, \citenamefont {Espinosa}, \citenamefont {Dral},\ and\ \citenamefont {Kananenka}}]{rodriguez22}%
  \BibitemOpen
  \bibfield  {author} {\bibinfo {author} {\bibfnamefont {L.~E.~H.}\ \bibnamefont {Rodríguez}}, \bibinfo {author} {\bibfnamefont {A.}~\bibnamefont {Ullah}}, \bibinfo {author} {\bibfnamefont {K.~J.~R.}\ \bibnamefont {Espinosa}}, \bibinfo {author} {\bibfnamefont {P.~O.}\ \bibnamefont {Dral}}, \ and\ \bibinfo {author} {\bibfnamefont {A.~A.}\ \bibnamefont {Kananenka}},\ }\bibfield  {title} {\enquote {\bibinfo {title} {A comparative study of different machine learning methods for dissipative quantum dynamics},}\ }\href {\doibase 10.1088/2632-2153/ac9a9d} {\bibfield  {journal} {\bibinfo  {journal} {Machine Learning: Science and Technology}\ }\textbf {\bibinfo {volume} {3}},\ \bibinfo {pages} {045016} (\bibinfo {year} {2022})}\BibitemShut {NoStop}%
\bibitem [{\citenamefont {Ullah}\ \emph {et~al.}(2024)\citenamefont {Ullah}, \citenamefont {Huang}, \citenamefont {Yang},\ and\ \citenamefont {Dral}}]{D4DD00153B}%
  \BibitemOpen
  \bibfield  {author} {\bibinfo {author} {\bibfnamefont {A.}~\bibnamefont {Ullah}}, \bibinfo {author} {\bibfnamefont {Y.}~\bibnamefont {Huang}}, \bibinfo {author} {\bibfnamefont {M.}~\bibnamefont {Yang}}, \ and\ \bibinfo {author} {\bibfnamefont {P.~O.}\ \bibnamefont {Dral}},\ }\bibfield  {title} {\enquote {\bibinfo {title} {Physics-informed neural networks and beyond: enforcing physical constraints in quantum dissipative dynamics},}\ }\href {\doibase 10.1039/D4DD00153B} {\bibfield  {journal} {\bibinfo  {journal} {Digital Discovery}\ ,\ \bibinfo {pages} {--}} (\bibinfo {year} {2024})}\BibitemShut {NoStop}%
\bibitem [{\citenamefont {Vaswani}\ \emph {et~al.}(2017)\citenamefont {Vaswani}, \citenamefont {Shazeer}, \citenamefont {Parmar}, \citenamefont {Uszkoreit}, \citenamefont {Jones}, \citenamefont {Gomez}, \citenamefont {Kaiser},\ and\ \citenamefont {Polosukhin}}]{vaswani2017attention}%
  \BibitemOpen
  \bibfield  {author} {\bibinfo {author} {\bibfnamefont {A.}~\bibnamefont {Vaswani}}, \bibinfo {author} {\bibfnamefont {N.}~\bibnamefont {Shazeer}}, \bibinfo {author} {\bibfnamefont {N.}~\bibnamefont {Parmar}}, \bibinfo {author} {\bibfnamefont {J.}~\bibnamefont {Uszkoreit}}, \bibinfo {author} {\bibfnamefont {L.}~\bibnamefont {Jones}}, \bibinfo {author} {\bibfnamefont {A.~N.}\ \bibnamefont {Gomez}}, \bibinfo {author} {\bibfnamefont {{\L}.}~\bibnamefont {Kaiser}}, \ and\ \bibinfo {author} {\bibfnamefont {I.}~\bibnamefont {Polosukhin}},\ }\bibfield  {title} {\enquote {\bibinfo {title} {Attention is all you need},}\ }\href@noop {} {\bibfield  {journal} {\bibinfo  {journal} {Advances in neural information processing systems}\ }\textbf {\bibinfo {volume} {30}} (\bibinfo {year} {2017})}\BibitemShut {NoStop}%
\bibitem [{\citenamefont {Dosovitskiy}\ \emph {et~al.}(2020)\citenamefont {Dosovitskiy}, \citenamefont {Beyer}, \citenamefont {Kolesnikov}, \citenamefont {Weissenborn}, \citenamefont {Zhai}, \citenamefont {Unterthiner}, \citenamefont {Dehghani}, \citenamefont {Minderer}, \citenamefont {Heigold}, \citenamefont {Gelly} \emph {et~al.}}]{dosovitskiy2020image}%
  \BibitemOpen
  \bibfield  {author} {\bibinfo {author} {\bibfnamefont {A.}~\bibnamefont {Dosovitskiy}}, \bibinfo {author} {\bibfnamefont {L.}~\bibnamefont {Beyer}}, \bibinfo {author} {\bibfnamefont {A.}~\bibnamefont {Kolesnikov}}, \bibinfo {author} {\bibfnamefont {D.}~\bibnamefont {Weissenborn}}, \bibinfo {author} {\bibfnamefont {X.}~\bibnamefont {Zhai}}, \bibinfo {author} {\bibfnamefont {T.}~\bibnamefont {Unterthiner}}, \bibinfo {author} {\bibfnamefont {M.}~\bibnamefont {Dehghani}}, \bibinfo {author} {\bibfnamefont {M.}~\bibnamefont {Minderer}}, \bibinfo {author} {\bibfnamefont {G.}~\bibnamefont {Heigold}}, \bibinfo {author} {\bibfnamefont {S.}~\bibnamefont {Gelly}},  \emph {et~al.},\ }\bibfield  {title} {\enquote {\bibinfo {title} {An image is worth 16x16 words: Transformers for image recognition at scale},}\ }\href@noop {} {\bibfield  {journal} {\bibinfo  {journal} {arXiv preprint arXiv:2010.11929}\ } (\bibinfo {year} {2020})}\BibitemShut {NoStop}%
\bibitem [{\citenamefont {Radford}\ \emph {et~al.}(2023)\citenamefont {Radford}, \citenamefont {Kim}, \citenamefont {Xu}, \citenamefont {Brockman}, \citenamefont {McLeavey},\ and\ \citenamefont {Sutskever}}]{radford2023robust}%
  \BibitemOpen
  \bibfield  {author} {\bibinfo {author} {\bibfnamefont {A.}~\bibnamefont {Radford}}, \bibinfo {author} {\bibfnamefont {J.~W.}\ \bibnamefont {Kim}}, \bibinfo {author} {\bibfnamefont {T.}~\bibnamefont {Xu}}, \bibinfo {author} {\bibfnamefont {G.}~\bibnamefont {Brockman}}, \bibinfo {author} {\bibfnamefont {C.}~\bibnamefont {McLeavey}}, \ and\ \bibinfo {author} {\bibfnamefont {I.}~\bibnamefont {Sutskever}},\ }\bibfield  {title} {\enquote {\bibinfo {title} {Robust speech recognition via large-scale weak supervision},}\ }in\ \href@noop {} {\emph {\bibinfo {booktitle} {International Conference on Machine Learning}}}\ (\bibinfo {organization} {PMLR},\ \bibinfo {year} {2023})\ pp.\ \bibinfo {pages} {28492--28518}\BibitemShut {NoStop}%
\bibitem [{\citenamefont {Wolf}\ \emph {et~al.}(2020)\citenamefont {Wolf}, \citenamefont {Debut}, \citenamefont {Sanh}, \citenamefont {Chaumond}, \citenamefont {Delangue}, \citenamefont {Moi}, \citenamefont {Cistac}, \citenamefont {Rault}, \citenamefont {Louf}, \citenamefont {Funtowicz} \emph {et~al.}}]{wolf2020transformers}%
  \BibitemOpen
  \bibfield  {author} {\bibinfo {author} {\bibfnamefont {T.}~\bibnamefont {Wolf}}, \bibinfo {author} {\bibfnamefont {L.}~\bibnamefont {Debut}}, \bibinfo {author} {\bibfnamefont {V.}~\bibnamefont {Sanh}}, \bibinfo {author} {\bibfnamefont {J.}~\bibnamefont {Chaumond}}, \bibinfo {author} {\bibfnamefont {C.}~\bibnamefont {Delangue}}, \bibinfo {author} {\bibfnamefont {A.}~\bibnamefont {Moi}}, \bibinfo {author} {\bibfnamefont {P.}~\bibnamefont {Cistac}}, \bibinfo {author} {\bibfnamefont {T.}~\bibnamefont {Rault}}, \bibinfo {author} {\bibfnamefont {R.}~\bibnamefont {Louf}}, \bibinfo {author} {\bibfnamefont {M.}~\bibnamefont {Funtowicz}},  \emph {et~al.},\ }\bibfield  {title} {\enquote {\bibinfo {title} {Transformers: State-of-the-art natural language processing},}\ }in\ \href@noop {} {\emph {\bibinfo {booktitle} {Proceedings of the 2020 conference on empirical methods in natural language processing: system demonstrations}}}\ (\bibinfo {year} {2020})\ pp.\ \bibinfo {pages} {38--45}\BibitemShut {NoStop}%
\bibitem [{\citenamefont {Donoso-Oliva}\ \emph {et~al.}(2023)\citenamefont {Donoso-Oliva}, \citenamefont {Becker}, \citenamefont {Protopapas}, \citenamefont {Cabrera-Vives}, \citenamefont {Vishnu},\ and\ \citenamefont {Vardhan}}]{donoso2023astromer}%
  \BibitemOpen
  \bibfield  {author} {\bibinfo {author} {\bibfnamefont {C.}~\bibnamefont {Donoso-Oliva}}, \bibinfo {author} {\bibfnamefont {I.}~\bibnamefont {Becker}}, \bibinfo {author} {\bibfnamefont {P.}~\bibnamefont {Protopapas}}, \bibinfo {author} {\bibfnamefont {G.}~\bibnamefont {Cabrera-Vives}}, \bibinfo {author} {\bibfnamefont {M.}~\bibnamefont {Vishnu}}, \ and\ \bibinfo {author} {\bibfnamefont {H.}~\bibnamefont {Vardhan}},\ }\bibfield  {title} {\enquote {\bibinfo {title} {Astromer-a transformer-based embedding for the representation of light curves},}\ }\href@noop {} {\bibfield  {journal} {\bibinfo  {journal} {Astronomy \& Astrophysics}\ }\textbf {\bibinfo {volume} {670}},\ \bibinfo {pages} {A54} (\bibinfo {year} {2023})}\BibitemShut {NoStop}%
\bibitem [{\citenamefont {Leggett}\ \emph {et~al.}(1987)\citenamefont {Leggett}, \citenamefont {Chakravarty}, \citenamefont {Dorsey}, \citenamefont {Fisher}, \citenamefont {Garg},\ and\ \citenamefont {Zwerger}}]{leggett1987dynamics}%
  \BibitemOpen
  \bibfield  {author} {\bibinfo {author} {\bibfnamefont {A.~J.}\ \bibnamefont {Leggett}}, \bibinfo {author} {\bibfnamefont {S.}~\bibnamefont {Chakravarty}}, \bibinfo {author} {\bibfnamefont {A.~T.}\ \bibnamefont {Dorsey}}, \bibinfo {author} {\bibfnamefont {M.~P.}\ \bibnamefont {Fisher}}, \bibinfo {author} {\bibfnamefont {A.}~\bibnamefont {Garg}}, \ and\ \bibinfo {author} {\bibfnamefont {W.}~\bibnamefont {Zwerger}},\ }\bibfield  {title} {\enquote {\bibinfo {title} {Dynamics of the dissipative two-state system},}\ }\href@noop {} {\bibfield  {journal} {\bibinfo  {journal} {Reviews of Modern Physics}\ }\textbf {\bibinfo {volume} {59}},\ \bibinfo {pages} {1} (\bibinfo {year} {1987})}\BibitemShut {NoStop}%
\bibitem [{\citenamefont {Makhlin}, \citenamefont {Sch{\"o}n},\ and\ \citenamefont {Shnirman}(2001)}]{makhlin2001quantum}%
  \BibitemOpen
  \bibfield  {author} {\bibinfo {author} {\bibfnamefont {Y.}~\bibnamefont {Makhlin}}, \bibinfo {author} {\bibfnamefont {G.}~\bibnamefont {Sch{\"o}n}}, \ and\ \bibinfo {author} {\bibfnamefont {A.}~\bibnamefont {Shnirman}},\ }\bibfield  {title} {\enquote {\bibinfo {title} {Quantum-state engineering with josephson-junction devices},}\ }\href@noop {} {\bibfield  {journal} {\bibinfo  {journal} {Reviews of modern physics}\ }\textbf {\bibinfo {volume} {73}},\ \bibinfo {pages} {357} (\bibinfo {year} {2001})}\BibitemShut {NoStop}%
\bibitem [{\citenamefont {Winter}\ \emph {et~al.}(2008)\citenamefont {Winter}, \citenamefont {Rieger}, \citenamefont {Vojta},\ and\ \citenamefont {Bulla}}]{winter2008quantum}%
  \BibitemOpen
  \bibfield  {author} {\bibinfo {author} {\bibfnamefont {A.}~\bibnamefont {Winter}}, \bibinfo {author} {\bibfnamefont {H.}~\bibnamefont {Rieger}}, \bibinfo {author} {\bibfnamefont {M.}~\bibnamefont {Vojta}}, \ and\ \bibinfo {author} {\bibfnamefont {R.}~\bibnamefont {Bulla}},\ }\bibfield  {title} {\enquote {\bibinfo {title} {The quantum phase transition in the sub-ohmic spin-boson model: Quantum monte-carlo study with a continuous imaginary time cluster algorithm},}\ }\href@noop {} {\bibfield  {journal} {\bibinfo  {journal} {arXiv preprint arXiv:0807.4716}\ } (\bibinfo {year} {2008})}\BibitemShut {NoStop}%
\bibitem [{\citenamefont {Alvermann}\ and\ \citenamefont {Fehske}(2008)}]{alvermann2008sparse}%
  \BibitemOpen
  \bibfield  {author} {\bibinfo {author} {\bibfnamefont {A.}~\bibnamefont {Alvermann}}\ and\ \bibinfo {author} {\bibfnamefont {H.}~\bibnamefont {Fehske}},\ }\bibfield  {title} {\enquote {\bibinfo {title} {Sparse polynomial space approach to dissipative quantum systems: Application to the sub-ohmic spin-boson model},}\ }\href@noop {} {\bibfield  {journal} {\bibinfo  {journal} {arXiv preprint arXiv:0812.2808}\ } (\bibinfo {year} {2008})}\BibitemShut {NoStop}%
\bibitem [{\citenamefont {Garg}, \citenamefont {Onuchic},\ and\ \citenamefont {Ambegaokar}(1985)}]{garg1985effect}%
  \BibitemOpen
  \bibfield  {author} {\bibinfo {author} {\bibfnamefont {A.}~\bibnamefont {Garg}}, \bibinfo {author} {\bibfnamefont {J.~N.}\ \bibnamefont {Onuchic}}, \ and\ \bibinfo {author} {\bibfnamefont {V.}~\bibnamefont {Ambegaokar}},\ }\bibfield  {title} {\enquote {\bibinfo {title} {Effect of friction on electron transfer in biomolecules},}\ }\href@noop {} {\bibfield  {journal} {\bibinfo  {journal} {The Journal of chemical physics}\ }\textbf {\bibinfo {volume} {83}},\ \bibinfo {pages} {4491--4503} (\bibinfo {year} {1985})}\BibitemShut {NoStop}%
\bibitem [{\citenamefont {Wang}\ \emph {et~al.}(1999)\citenamefont {Wang}, \citenamefont {Song}, \citenamefont {Chandler},\ and\ \citenamefont {Miller}}]{wang1999semiclassical}%
  \BibitemOpen
  \bibfield  {author} {\bibinfo {author} {\bibfnamefont {H.}~\bibnamefont {Wang}}, \bibinfo {author} {\bibfnamefont {X.}~\bibnamefont {Song}}, \bibinfo {author} {\bibfnamefont {D.}~\bibnamefont {Chandler}}, \ and\ \bibinfo {author} {\bibfnamefont {W.~H.}\ \bibnamefont {Miller}},\ }\bibfield  {title} {\enquote {\bibinfo {title} {Semiclassical study of electronically nonadiabatic dynamics in the condensed-phase: Spin-boson problem with debye spectral density},}\ }\href@noop {} {\bibfield  {journal} {\bibinfo  {journal} {The Journal of chemical physics}\ }\textbf {\bibinfo {volume} {110}},\ \bibinfo {pages} {4828--4840} (\bibinfo {year} {1999})}\BibitemShut {NoStop}%
\bibitem [{\citenamefont {Niu}, \citenamefont {Zhong},\ and\ \citenamefont {Yu}(2021)}]{niu2021review}%
  \BibitemOpen
  \bibfield  {author} {\bibinfo {author} {\bibfnamefont {Z.}~\bibnamefont {Niu}}, \bibinfo {author} {\bibfnamefont {G.}~\bibnamefont {Zhong}}, \ and\ \bibinfo {author} {\bibfnamefont {H.}~\bibnamefont {Yu}},\ }\bibfield  {title} {\enquote {\bibinfo {title} {A review on the attention mechanism of deep learning},}\ }\href@noop {} {\bibfield  {journal} {\bibinfo  {journal} {Neurocomputing}\ }\textbf {\bibinfo {volume} {452}},\ \bibinfo {pages} {48--62} (\bibinfo {year} {2021})}\BibitemShut {NoStop}%
\bibitem [{\citenamefont {Corbetta}\ and\ \citenamefont {Shulman}(2002)}]{corbetta2002control}%
  \BibitemOpen
  \bibfield  {author} {\bibinfo {author} {\bibfnamefont {M.}~\bibnamefont {Corbetta}}\ and\ \bibinfo {author} {\bibfnamefont {G.~L.}\ \bibnamefont {Shulman}},\ }\bibfield  {title} {\enquote {\bibinfo {title} {Control of goal-directed and stimulus-driven attention in the brain},}\ }\href@noop {} {\bibfield  {journal} {\bibinfo  {journal} {Nature reviews neuroscience}\ }\textbf {\bibinfo {volume} {3}},\ \bibinfo {pages} {201--215} (\bibinfo {year} {2002})}\BibitemShut {NoStop}%
\bibitem [{\citenamefont {Bahdanau}, \citenamefont {Cho},\ and\ \citenamefont {Bengio}(2014)}]{bahdanau2014neural}%
  \BibitemOpen
  \bibfield  {author} {\bibinfo {author} {\bibfnamefont {D.}~\bibnamefont {Bahdanau}}, \bibinfo {author} {\bibfnamefont {K.}~\bibnamefont {Cho}}, \ and\ \bibinfo {author} {\bibfnamefont {Y.}~\bibnamefont {Bengio}},\ }\bibfield  {title} {\enquote {\bibinfo {title} {Neural machine translation by jointly learning to align and translate},}\ }\href@noop {} {\bibfield  {journal} {\bibinfo  {journal} {arXiv preprint arXiv:1409.0473}\ } (\bibinfo {year} {2014})}\BibitemShut {NoStop}%
\bibitem [{\citenamefont {Aliabadi}\ \emph {et~al.}(2020)\citenamefont {Aliabadi}, \citenamefont {Emami}, \citenamefont {Dong},\ and\ \citenamefont {Huang}}]{aliabadi2020attention}%
  \BibitemOpen
  \bibfield  {author} {\bibinfo {author} {\bibfnamefont {M.~M.}\ \bibnamefont {Aliabadi}}, \bibinfo {author} {\bibfnamefont {H.}~\bibnamefont {Emami}}, \bibinfo {author} {\bibfnamefont {M.}~\bibnamefont {Dong}}, \ and\ \bibinfo {author} {\bibfnamefont {Y.}~\bibnamefont {Huang}},\ }\bibfield  {title} {\enquote {\bibinfo {title} {Attention-based recurrent neural network for multistep-ahead prediction of process performance},}\ }\href@noop {} {\bibfield  {journal} {\bibinfo  {journal} {Computers \& Chemical Engineering}\ }\textbf {\bibinfo {volume} {140}},\ \bibinfo {pages} {106931} (\bibinfo {year} {2020})}\BibitemShut {NoStop}%
\bibitem [{\citenamefont {Moreno-Cartagena}\ \emph {et~al.}(2023)\citenamefont {Moreno-Cartagena}, \citenamefont {Cabrera-Vives}, \citenamefont {Protopapas}, \citenamefont {Donoso-Oliva}, \citenamefont {P{\'e}rez-Carrasco},\ and\ \citenamefont {C{\'a}diz-Leyton}}]{moreno2023positional}%
  \BibitemOpen
  \bibfield  {author} {\bibinfo {author} {\bibfnamefont {D.}~\bibnamefont {Moreno-Cartagena}}, \bibinfo {author} {\bibfnamefont {G.}~\bibnamefont {Cabrera-Vives}}, \bibinfo {author} {\bibfnamefont {P.}~\bibnamefont {Protopapas}}, \bibinfo {author} {\bibfnamefont {C.}~\bibnamefont {Donoso-Oliva}}, \bibinfo {author} {\bibfnamefont {M.}~\bibnamefont {P{\'e}rez-Carrasco}}, \ and\ \bibinfo {author} {\bibfnamefont {M.}~\bibnamefont {C{\'a}diz-Leyton}},\ }\bibfield  {title} {\enquote {\bibinfo {title} {Positional encodings for light curve transformers: Playing with positions and attention},}\ }\href@noop {} {\bibfield  {journal} {\bibinfo  {journal} {arXiv preprint arXiv:2308.06404}\ } (\bibinfo {year} {2023})}\BibitemShut {NoStop}%
\bibitem [{\citenamefont {Johansson}, \citenamefont {Nation},\ and\ \citenamefont {Nori}(2012)}]{johansson2012qutip}%
  \BibitemOpen
  \bibfield  {author} {\bibinfo {author} {\bibfnamefont {J.~R.}\ \bibnamefont {Johansson}}, \bibinfo {author} {\bibfnamefont {P.~D.}\ \bibnamefont {Nation}}, \ and\ \bibinfo {author} {\bibfnamefont {F.}~\bibnamefont {Nori}},\ }\bibfield  {title} {\enquote {\bibinfo {title} {Qutip: An open-source python framework for the dynamics of open quantum systems},}\ }\href@noop {} {\bibfield  {journal} {\bibinfo  {journal} {Computer physics communications}\ }\textbf {\bibinfo {volume} {183}},\ \bibinfo {pages} {1760--1772} (\bibinfo {year} {2012})}\BibitemShut {NoStop}%
\bibitem [{\citenamefont {Chollet}\ \emph {et~al.}(2015)\citenamefont {Chollet} \emph {et~al.}}]{keras}%
  \BibitemOpen
  \bibfield  {author} {\bibinfo {author} {\bibfnamefont {F.}~\bibnamefont {Chollet}} \emph {et~al.},\ }\href@noop {} {\enquote {\bibinfo {title} {Keras},}\ } (\bibinfo {year} {2015}),\ \bibinfo {note} {\url{https://github.com/fchollet/keras}}\BibitemShut {NoStop}%
\bibitem [{\citenamefont {Abadi}\ \emph {et~al.}(2015)\citenamefont {Abadi}, \citenamefont {Agarwal}, \citenamefont {Barham}, \citenamefont {Brevdo}, \citenamefont {Chen}, \citenamefont {Citro}, \citenamefont {Corrado}, \citenamefont {Davis}, \citenamefont {Dean}, \citenamefont {Devin}, \citenamefont {Ghemawat}, \citenamefont {Goodfellow}, \citenamefont {Harp}, \citenamefont {Irving}, \citenamefont {Isard}, \citenamefont {Jia}, \citenamefont {Jozefowicz}, \citenamefont {Kaiser}, \citenamefont {Kudlur}, \citenamefont {Levenberg}, \citenamefont {Man\'{e}}, \citenamefont {Monga}, \citenamefont {Moore}, \citenamefont {Murray}, \citenamefont {Olah}, \citenamefont {Schuster}, \citenamefont {Shlens}, \citenamefont {Steiner}, \citenamefont {Sutskever}, \citenamefont {Talwar}, \citenamefont {Tucker}, \citenamefont {Vanhoucke}, \citenamefont {Vasudevan}, \citenamefont {Vi\'{e}gas}, \citenamefont {Vinyals}, \citenamefont {Warden}, \citenamefont {Wattenberg}, \citenamefont {Wicke}, \citenamefont {Yu},\ and\ \citenamefont
  {Zheng}}]{tensorflow}%
  \BibitemOpen
  \bibfield  {author} {\bibinfo {author} {\bibfnamefont {M.}~\bibnamefont {Abadi}}, \bibinfo {author} {\bibfnamefont {A.}~\bibnamefont {Agarwal}}, \bibinfo {author} {\bibfnamefont {P.}~\bibnamefont {Barham}}, \bibinfo {author} {\bibfnamefont {E.}~\bibnamefont {Brevdo}}, \bibinfo {author} {\bibfnamefont {Z.}~\bibnamefont {Chen}}, \bibinfo {author} {\bibfnamefont {C.}~\bibnamefont {Citro}}, \bibinfo {author} {\bibfnamefont {G.~S.}\ \bibnamefont {Corrado}}, \bibinfo {author} {\bibfnamefont {A.}~\bibnamefont {Davis}}, \bibinfo {author} {\bibfnamefont {J.}~\bibnamefont {Dean}}, \bibinfo {author} {\bibfnamefont {M.}~\bibnamefont {Devin}}, \bibinfo {author} {\bibfnamefont {S.}~\bibnamefont {Ghemawat}}, \bibinfo {author} {\bibfnamefont {I.}~\bibnamefont {Goodfellow}}, \bibinfo {author} {\bibfnamefont {A.}~\bibnamefont {Harp}}, \bibinfo {author} {\bibfnamefont {G.}~\bibnamefont {Irving}}, \bibinfo {author} {\bibfnamefont {M.}~\bibnamefont {Isard}}, \bibinfo {author} {\bibfnamefont {Y.}~\bibnamefont {Jia}}, \bibinfo
  {author} {\bibfnamefont {R.}~\bibnamefont {Jozefowicz}}, \bibinfo {author} {\bibfnamefont {L.}~\bibnamefont {Kaiser}}, \bibinfo {author} {\bibfnamefont {M.}~\bibnamefont {Kudlur}}, \bibinfo {author} {\bibfnamefont {J.}~\bibnamefont {Levenberg}}, \bibinfo {author} {\bibfnamefont {D.}~\bibnamefont {Man\'{e}}}, \bibinfo {author} {\bibfnamefont {R.}~\bibnamefont {Monga}}, \bibinfo {author} {\bibfnamefont {S.}~\bibnamefont {Moore}}, \bibinfo {author} {\bibfnamefont {D.}~\bibnamefont {Murray}}, \bibinfo {author} {\bibfnamefont {C.}~\bibnamefont {Olah}}, \bibinfo {author} {\bibfnamefont {M.}~\bibnamefont {Schuster}}, \bibinfo {author} {\bibfnamefont {J.}~\bibnamefont {Shlens}}, \bibinfo {author} {\bibfnamefont {B.}~\bibnamefont {Steiner}}, \bibinfo {author} {\bibfnamefont {I.}~\bibnamefont {Sutskever}}, \bibinfo {author} {\bibfnamefont {K.}~\bibnamefont {Talwar}}, \bibinfo {author} {\bibfnamefont {P.}~\bibnamefont {Tucker}}, \bibinfo {author} {\bibfnamefont {V.}~\bibnamefont {Vanhoucke}}, \bibinfo {author}
  {\bibfnamefont {V.}~\bibnamefont {Vasudevan}}, \bibinfo {author} {\bibfnamefont {F.}~\bibnamefont {Vi\'{e}gas}}, \bibinfo {author} {\bibfnamefont {O.}~\bibnamefont {Vinyals}}, \bibinfo {author} {\bibfnamefont {P.}~\bibnamefont {Warden}}, \bibinfo {author} {\bibfnamefont {M.}~\bibnamefont {Wattenberg}}, \bibinfo {author} {\bibfnamefont {M.}~\bibnamefont {Wicke}}, \bibinfo {author} {\bibfnamefont {Y.}~\bibnamefont {Yu}}, \ and\ \bibinfo {author} {\bibfnamefont {X.}~\bibnamefont {Zheng}},\ }\href {https://www.tensorflow.org/} {\enquote {\bibinfo {title} {{TensorFlow}: Large-scale machine learning on heterogeneous systems},}\ } (\bibinfo {year} {2015}),\ \bibinfo {note} {software available from tensorflow.org}\BibitemShut {NoStop}%
\bibitem [{\citenamefont {Diederik}(2014)}]{diederik2014adam}%
  \BibitemOpen
  \bibfield  {author} {\bibinfo {author} {\bibfnamefont {P.~K.}\ \bibnamefont {Diederik}},\ }\bibfield  {title} {\enquote {\bibinfo {title} {Adam: A method for stochastic optimization},}\ }\href@noop {} {\bibfield  {journal} {\bibinfo  {journal} {(No Title)}\ } (\bibinfo {year} {2014})}\BibitemShut {NoStop}%
\bibitem [{\citenamefont {O'Malley}\ \emph {et~al.}(2019)\citenamefont {O'Malley}, \citenamefont {Bursztein}, \citenamefont {Long}, \citenamefont {Chollet}, \citenamefont {Jin}, \citenamefont {Invernizzi} \emph {et~al.}}]{omalley2019kerastuner}%
  \BibitemOpen
  \bibfield  {author} {\bibinfo {author} {\bibfnamefont {T.}~\bibnamefont {O'Malley}}, \bibinfo {author} {\bibfnamefont {E.}~\bibnamefont {Bursztein}}, \bibinfo {author} {\bibfnamefont {J.}~\bibnamefont {Long}}, \bibinfo {author} {\bibfnamefont {F.}~\bibnamefont {Chollet}}, \bibinfo {author} {\bibfnamefont {H.}~\bibnamefont {Jin}}, \bibinfo {author} {\bibfnamefont {L.}~\bibnamefont {Invernizzi}},  \emph {et~al.},\ }\href@noop {} {\enquote {\bibinfo {title} {Kerastuner},}\ }\bibinfo {howpublished} {\url{https://github.com/keras-team/keras-tuner}} (\bibinfo {year} {2019})\BibitemShut {NoStop}%
\end{thebibliography}%

\end{document}